%% file: DiffusionBiomolecules.tex
\documentclass{article}
\usepackage[margin=1in]{geometry}

\usepackage{caption}
\usepackage{subcaption}
\usepackage{graphicx} % Required for inserting images

\usepackage{enumitem}
\usepackage{amsmath,amsfonts}
\usepackage[]{bm}
\usepackage{multirow}
\usepackage[]{hyperref}
\usepackage{cleveref}
\usepackage{adjustbox}
\usepackage{booktabs}
\usepackage[]{tikz}

\usepackage{authblk}

\usepackage{ifthen}
\newboolean{isJMB}
\setboolean{isJMB}{false}

\title{Sifting through the Noise: A Survey of Diffusion Probabilistic Models and Their Applications to Biomolecules}
\author[1]{Trevor Norton}
\author[1]{Debswapna Bhattacharya\thanks{Correspondence to \href{mailto:dbhattacharya@vt.edu}{dbhattacharya@vt.edu}}}
\affil[1]{Department of Computer Science, Virginia Tech, Blacksburg, Virginia, 24061, USA}
\date{}

\newcommand{\bx}[1]{\mathbf{x}_{#1}}

\newcommand{\mcN}{\mathcal{N}}
\newcommand{\rd}{\mathrm{d}}

\newcommand{\se}[1]{\mathrm{SE}(#1)}
\newcommand{\so}[1]{\mathrm{SO}(#1)}
\newcommand{\igso}{\mathrm{IGSO}(3)}

\begin{document}
	
	\maketitle

\input{sections/abstract.tex}
	
\input{sections/introduction.tex}

\input{sections/preliminaries.tex}

\input{sections/motifs.tex}

\input{sections/applications.tex}

\input{sections/conclusion.tex}

\input{sections/acknowledgments.tex}

\bibliographystyle{unsrt}
\bibliography{DiffusionBiomolecules.bib}

\appendix

\input{sections/benchmarking.tex}

\end{document}

%% file: sections/abstract.tex
\begin{abstract}
	Diffusion probabilistic models have made their way into a number of high-profile applications since their inception. In particular, there has been a wave of research into using diffusion models in the prediction and design of biomolecular structures and sequences. Their growing ubiquity makes it imperative for researchers in these fields to understand them. This paper serves as a general overview for the theory behind these models and the current state of research. We first introduce diffusion models and discuss common motifs used when applying them to biomolecules. We then present the significant outcomes achieved through the application of these models in generative and predictive tasks. This survey aims to provide readers with a comprehensive understanding of the increasingly critical role of diffusion models.
\end{abstract}

%% file: sections/introduction.tex
\section{Introduction}

Diffusion probabilistic models (or more simply diffusion models) have attracted attentions from researchers and the public alike for their success in image generation. After usurping generative adversarial networks (GANs) in sampling high-quality images \cite{goodfellow2020generative,dhariwal_diffusion_2021}, we have seen many diffusion-based image synthesis models introduced. Diffusion models have also been successfully applied to problems in computer vision \cite{croitoru2023diffusion}, audio generation \cite{popov2021diffusion, wu2021otts,liu2022diffsinger}, robotics \cite{carvalho2022conditioned,carvalho2023motion,kapelyukh2023dall,urain2023se,findlay2022denoising}, and many others.

Diffusion models are a category of deep generative models, which seek to easily sample some underlying distribution \(p(x)\). Sampling a distribution can be difficult in cases where the distribution has many modes or is in a high-dimensional space. This problem can be especially prevalent when working with thermodynamic systems, where energy barriers can prevent adequate exploration of the distribution. Non-machine learning techniques have relied on expensive search methods, such as Markov chain Monte Carlo (MCMC) simulations, to accurately sample a distribution \cite{chaudhury2010pyrosetta,koes2013lessons,friesner2004glide,heilmann2020sampling}. Diffusion models instead sample a tractable prior (a normal distribution, for example), which is then transformed into the correct distribution. They do this by adding Gaussian noise to the data distribution and then learning how to iteratively remove the noise. This iterative denoising breaks down the generation process into simpler, learnable steps and allows sampling from very high-dimensional, rough distributions.

This ability makes diffusion models an attractive choice for the modeling of biomolecules. Problems in the field typically result in very high-dimensional data that have rough distributions. Levinthal's paradox \cite{levinthal1969fold} famously noted this difficulty with the protein folding problem: put simply, if one were to search every possible conformation of a protein for its native strucutre, it would take longer than the age of the universe (even if checking the state only took on the order of picoseconds). The outlook worsens when one considers that the energy landscape may have several meta-stable states that are not the native structure. Because of this, straight-forward calculations are usually intractable, and deep learning methods have provided tremendous leaps forward in predictive tasks. Most saliently, AlphaFold2 has been able to provide experimental accuracy of the native folded structures for many proteins \cite{jumper_highly_2021}. Diffusion models benefit from scalable deep learning architectures and can circumvent the issues of dimensionality through the iterative process. Furthermore, the diffusion process progressively smooths out the underlying distribution, which makes the new, smoother distributions amenable to approximation via deep neural networks \cite{lu2021deep}.

Research into applications of diffusion models to biomolecules has exploded in recent years, and there have already been a series of successes. This paper seeks to give an overview of the current state of diffusion models in the field and highlight the strengths and weaknesses of the method. We will first proceed by giving a brief history of diffusion models and their derivations. Following this, we details some of the most common techniques used when directly applying diffusion models to problems involving biomolecules. The remainder of the paper summarizes recent results and looks at possible future areas of research.

\begin{figure}[ht]
	\centering
	\includegraphics[width=\textwidth]{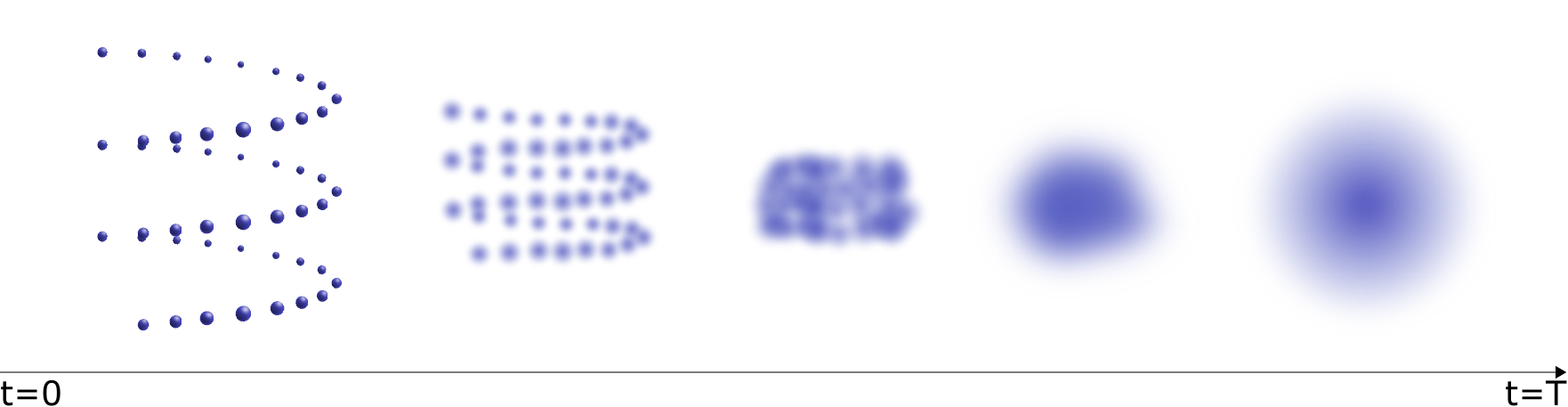}
	\caption{Illustration of a diffusion process in three dimensions. A molecule, such as a protein or RNA, can be represented by a collection of points in Euclidean space. The diffusion process gradually adds noise until the distribution is approximately normal. A diffusion model learns to denoise, so that samples from a normal distribution can be converted back to samples of the original distribution.}
\end{figure}

%% file: sections/preliminaries.tex
\section{Preliminaries on diffusion models}

Early diffusion models were heavily inspired by techniques from thermodynamics and statistics. The common observation among them was that it is possible to progressively transform a normal distribution into a given data distribution. However, the theories behind this transformation differed. Some viewed this process as learning to reverse a noising process, where the model progressively removes noise from a sample. Others thought about this process as accurately sampling a continuous family of perturbed distributions. Both theories can be unified into a flexible framework where the diffusion is described by continuous processes. In this section, we briefly recall the early work on diffusion models and the main results.

\subsection{Denoising diffusion probabilistic modeling}

Inspired by techniques in nonequlibrium thermodynamics and statistics \cite{neal2001annealed,jarzynski1997equilibrium}, Sohl-Dickstein et al.\ first proposed diffusion models as a way to develop probabilistic models that are flexible enough to capture complex distributions while providing exact sampling \cite{sohl-dickstein_deep_2015}. The main idea behind the algorithm was to start with samples from the desired distribution and iterative inject Gaussian noise until approximately reaching a stationary standard normal distribution. Training the model then comes down to learning how to reverse the noising process and recover the initial distribution.

More concretely, we want to approximate the distribution of data given as \(\bx 0 \sim p_{\text{data}}(\bx 0 )\). The forward process is a Markov chain whose transitions are given by Gaussian distributions according to a variance schedule \(0<\beta_1, \ldots, \beta_T<1\):
\begin{equation*}
	q(\bx t \mid \bx {t-1}) := \mcN (\bx t ; \sqrt{1-\beta_t}\bx{t-1}, \beta_t \mathbf I).
\end{equation*}
The transition kernel can be given exactly as 
\begin{equation*}
	q (\bx t \mid \bx 0) = \mcN(\bx t ; \sqrt{\alpha_t}\bx 0, (1-\alpha_t) \mathbf I),
\end{equation*}
where \(\alpha_t = \Pi_{t=1}^t (1-\beta_t)\), and so the exact distribution at time \(t\) is
\begin{equation*}
	q(\bx t) = \int q(\bx t \mid \bx 0) p_{\text{data}}(\bx 0)\, \mathrm d \bx 0.
\end{equation*} 
After sufficiently many steps through the Markov chain, we have that \(q(\bx T)\) is approximately equal to  \(\mcN(\bx T \mathbf 0, \mathbf I)\), the stationary distribution of the process.

To recover \(p_{\text{data}}\), we want to parameterize a reverse Markov chain \(p_{\theta}\) that will take the distribution \(\mcN(\mathbf 0, \mathbf I)\) and iteratively remove the added Gaussian noise. If the size of the steps \(\beta_t\) are small enough, the reverse process can be approximated by Gaussian transitions \cite{sohl-dickstein_deep_2015} and so the reverse process is parameterized as
\begin{equation*}
	p_\theta(\bx {t-1} \mid \bx {t})  := \mcN(\bx {t-1}; {\bm\mu}_\theta(\bx t, t), \mathbf\Sigma_\theta(\bx t, t)).
\end{equation*}
Training in \cite{sohl-dickstein_deep_2015} is then based on maximizing an evidence lower bound (ELBO).

In \cite{ho_denoising_2020}, this model is updated to have a simplified variational bound which improves sample quality. Reparameterizing the reverse process by setting
\begin{equation*}
	\bm{\mu}_\theta(\bx t, t) = \frac 1 {\sqrt{1-\beta_t}}\left( \bx t  - \frac{\beta_t}{\sqrt{1 -\alpha_t}}\bm{\epsilon}_\theta(\bx t, t) \right)
\end{equation*}
and rescaling the loss gives a simplified training objective
\begin{equation*}
	L_{\text{DDPM}}(\theta) := \mathbb{E}_{t,\bx 0, \bm\epsilon}\left[ \| \bm\epsilon - \bm\epsilon_\theta(\sqrt{\alpha_t} \bx 0  + \sqrt{1 - \alpha_t} \bm\epsilon, t)\|^2 \right],
\end{equation*}
where \(\bm\epsilon \sim \mcN(\mathbf 0, \mathbf I)\). This diffusion model with \(L_{\text{DDPM}}\) is known as a Denoising Diffusion Probabilistic Model (DDPM).

\subsection{Noise conditional score network}

In \cite{song_generative_2019}, the authors approach the problem of sampling from an unknown distribution \(p_{\text{data}}\) by estimating its (Stein) score \(\nabla_{\mathbf x} \log p_{\text{data}}(\mathbf x)\) at varying noise levels. The score is related to ideas from statistical mechanics. Namely, if we have an energy function \(E(\mathbf x)\), then the associated Boltzmann distribution of the data is given according to 
\begin{equation*}
	p(\mathbf x)  = \frac{\exp(-E(\mathbf x))}{Z},
\end{equation*}
where \(Z\) is a normalizing factor that is typically intractable to compute. Thus determining the score of the distribution is essentially equivalent to finding \(-\nabla E(\mathbf x)\), from which Langevin dynamics of the system may be computed.

Theoretically, approximating \(\nabla_{\mathbf x} \log p_{\text{data}}\) would allow sampling of the distribution directly through simulations. However, as Song et al.\ note, there are two major challenges with this approach:
\begin{enumerate}[label=(\arabic*)]
	\item Typically data lies on a low-dimensional manifold in the ambient space, in which case the score may be undefined or difficult to accurately approximate.
	\item There may be low-density regions of the distribution. This results in poor initial sampling of the distribution in those spaces making it harder to estimate the score. Furthermore, when sampling with Langevin dynamics, slow mixing can occur leading inaccurate estimates of the distribution in the short term.
\end{enumerate}
Both challenges can be addressed by instead estimating the scores of perturbations of \(p_{\text{data}}\). The perturbation kernel is defined as 
\begin{equation*}
	q_\sigma(\mathbf x\mid \mathbf x') := \mcN(\mathbf x; \mathbf x', \sigma^2\mathbf I),
\end{equation*}
so that the data distribution is perturbed to 
\begin{equation*}
	q_\sigma(\mathbf x) = \int q(\mathbf x \mid \mathbf x') p_{\text{data}}(\mathbf x')\, \mathrm d \mathbf x'.
\end{equation*}
The aim of the model is to approximate the score \(\nabla_{\mathbf x} \log q_\sigma(\mathbf x)\) for increasing levels of noise \(\sigma_{\text{min}} = \sigma_1 < \sigma_2 < \cdots < \sigma_N = \sigma_{\text{max}}\). The levels of noise are chosen so that \(q_{\sigma_{\text{min}}}(\mathbf x) \approx p_{\text{data}}(\mathbf x)\) and \(q_{\sigma_{\text{max}}}(\mathbf x) \approx \mcN(\mathbf x; \mathbf 0, \mathbf I)\). Thus there will be a family of distributions ending with pure noise and progressively getting closer to the desired distribution.

A \emph{noise conditional score network} (NCSN) \(s_\theta(\mathbf x, \sigma)\) is trained to estimate the scores at the specified noise levels. There are several techniques to approximate the scores, but most implementations use \emph{denoising score matching} which seeks to minimize
\begin{equation*}
	\mathbb{E}_{q_\sigma(\mathbf x| \mathbf x') p_{\text{data}}(\mathbf x')}\left[\| s_\theta(\mathbf x, \sigma)  - \nabla_{\mathbf x}\log q_\sigma (\mathbf x | \mathbf x')\|^2\right].
\end{equation*}
Since the perturbation kernel is Gaussian, the gradient can be written exactly. Song et al.\ use the training objective
\begin{equation*}
	L_{\text{NCSN}} (\theta) := \sum_{i=1}^N \sigma_i^2 \mathbb{E}_{p_{\text{data}}(\mathbf x')} \mathbb{E}_{q_{\sigma_i}(\mathbf x\mid \mathbf x')} \left[ \left\| s_\theta (\mathbf x , \sigma_i) + \frac{\mathbf x -\mathbf x'}{\sigma_i^2}\right\|^2 \right]
\end{equation*}

After training and determining \(\theta^* = \mathrm{argmin}_\theta L_{\text{NCSN}}(\theta)\), sampling from the distribution is done by a series of Langevin simulations at each noise level. That is, \(M\) steps of the (overdamped) Langevin dynamics are performed for each  \(\sigma_i\):
\begin{equation*}
	\bx i ^ m = \bx i ^{m-1} + \varepsilon_i s_{\theta^*}(\bx i ^{m-1}, \sigma_i) + \sqrt{2\varepsilon_i} \mathbf{z}_i^m, \quad m = 1,2,\ldots, M
\end{equation*}
where \(\varepsilon_i> 0\) is the step size and \(\mathbf{z}_i^m\) is standard normal. Then the final sample at the noise level \(\mathbf{x}^M_i\) is approximately distributed as \(q_{\sigma_i}(\mathbf x)\). The sampling is initialized with \(\mathbf{x}^0_N \sim \mcN(\mathbf x \mid \mathbf 0, \sigma_{\text{max}}^2 \mathbf I)\). And each noise level is initialized with the terminal value of the previous noise level: \(\mathbf{x}^0_i = \mathbf{x}^M_{i+1}\). The final value, \(\mathbf x^M_1\) is an approximate sample from \(p_{\text{data}}(\mathbf x)\).
\subsection{Score based modeling with stochastic differential equations}

Previous diffusion models used discrete Gaussian noising process to transform their distributions into standard normal distributions. However, modeling the noising process as a stochastic differential equation (SDE), as suggested in \cite{song_score-based_2020}, has many advantages. In particular, the framework is flexible and generalizes DDPMs and NCSNs as discretizations of different equations.

The forward diffusion process is modeled with an Itô SDE:
\begin{equation}\label{eq:sde}
	\rd\mathbf x = \mathbf f (\mathbf x, t) \rd t + g(t) \rd \mathbf w,
\end{equation}
where \(\mathbf w\) is a standard Wiener process, \(\mathbf f : \mathbb R^d \times [0,\infty) \to \mathbb R^d\) is the \emph{drift} coefficient and \(g:[0,\infty) \to \mathbb R\) is the \emph{diffusion} coefficient. The initial condition \(\bx 0\) has the distribution of the data, \(p_0(\mathbf x)\), and it is assumed that for a sufficiently long time \(T\) that \(\bx T\) is approximately distributed as the stationary solution of \cref{eq:sde}.

Taking the step sizes for DDPMs and NCSNs to zero gives a continuous version of the models that we can write as SDEs. The DDPM converges to the SDE
\begin{equation*}
	\mathrm{d}\mathbf x  = - \frac 1 2 \beta(t) \mathbf x \mathrm d t + \sqrt{\beta(t)} \mathrm{d}\mathbf{w},
\end{equation*}
which we refer to as the variance preserving (VP) SDE. Also, NCSNs converge to the SDE
\begin{equation*}
	\mathrm d \mathbf x = \sqrt{\frac{\mathrm d [\sigma(t)^2]}{\mathrm d t}}\mathrm d \mathbf w,
\end{equation*}
which we refer to as the variance exploding (VE) SDE.

If the score \(\nabla_{\mathbf x} \log p_t(\mathbf x)\) is known, then the diffusion process can be reversed by the following SDE:
\begin{equation}\label{eq:reverse-sde}
	\rd \mathbf x = \left[ \mathbf f (\mathbf x , t) - g(t)^2 \nabla_{\mathbf x} \log p_t(\mathbf x) \right] \rd t + g(t) \rd \overline{\mathbf w},
\end{equation}
where \(\overline{\mathbf w}\) is a standard Wiener process when time flows backward from \(T\) to \(0\). Thus the goal is to train a score network \(s_\theta (\mathbf x, t)\) to approximate \(\nabla_{\mathbf x} \log p_t(\mathbf x)\). A typical objective for training is by again using denoising score matching:
\begin{equation*}
	\mathbb{E}_t \Big\{\lambda(t) \mathbb{E}_{p_0(\mathbf x')} \mathbb{E}_{p_{0,t}(\mathbf x\mid \mathbf x')} \left[\| s_\theta(\mathbf x, t) - \nabla_{\mathbf x} \log  p_{0,t}(\mathbf x\mid \mathbf x')\|^2\right]\Big\}.
\end{equation*}
Other score-matching objectives can also be used here, such as sliced score matching \cite{song2020sliced}.

Apart from \cref{eq:reverse-sde}, there is another method for reversing the diffusion process. The ODE
\begin{equation}\label{eq:ode}
	\rd \mathbf x = \left[ \mathbf f (\mathbf x, t) - \frac 1 2 g(t)^2 \nabla_{\mathbf x} \log p_t(\mathbf x)\right] \rd t
\end{equation}
shares the same marginal probability densities as the SDE. One benefit of \cref{eq:ode} is that sampling can be done much quicker using standard ODE solvers versus the Euler–Maruyama method for integrating \cref{eq:reverse-sde}. Furthermore, integrating the ODE provides identifiable latent representations of the data and exact likelihood computations.

\ifthenelse{\boolean{isJMB}}{
	% Formatting for JMB
	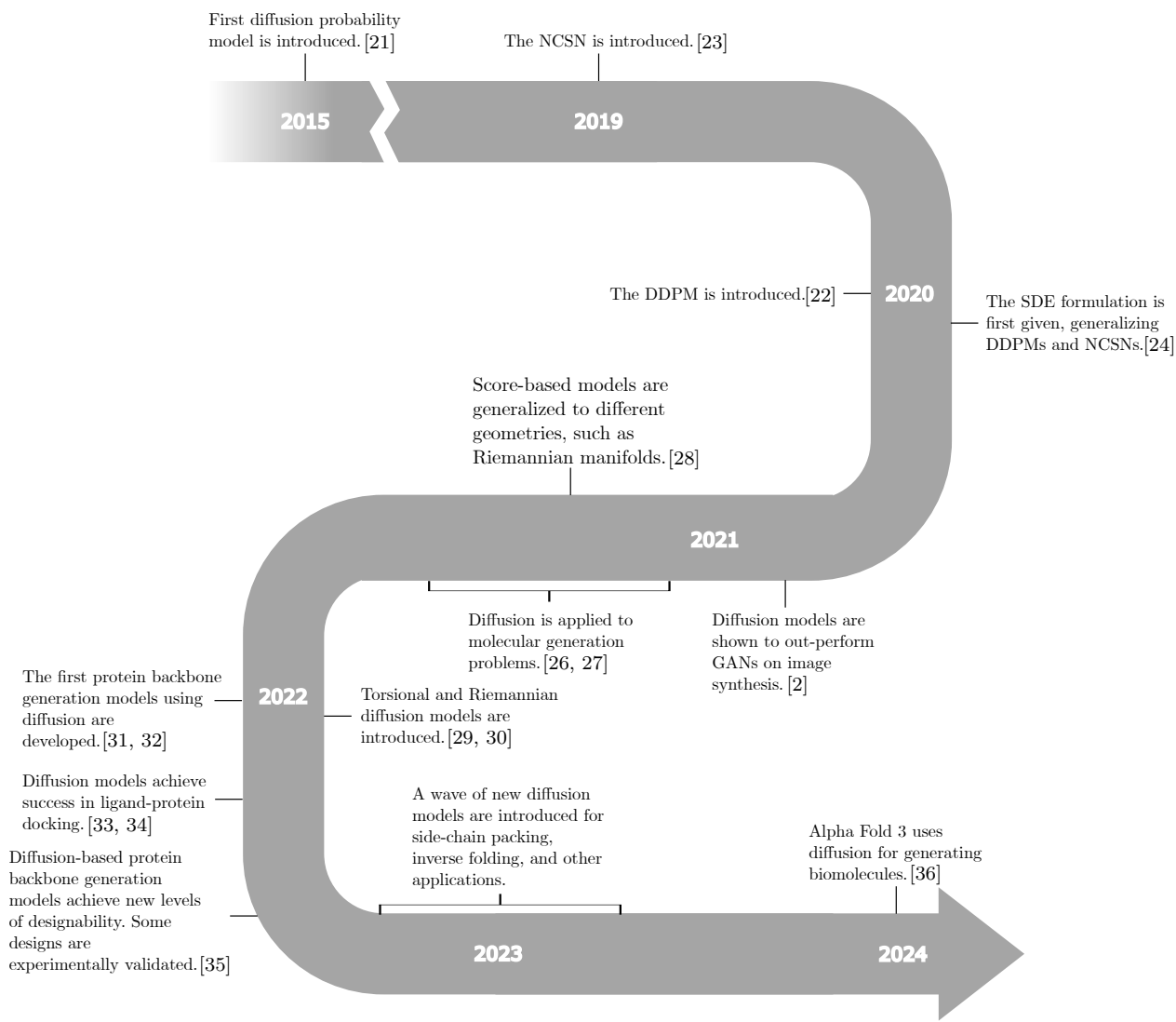
\begin{figure}[h]
		\begin{tikzpicture}
			\node[anchor=south west,inner sep=0] at (0,0) {\includegraphics[width=\textwidth]{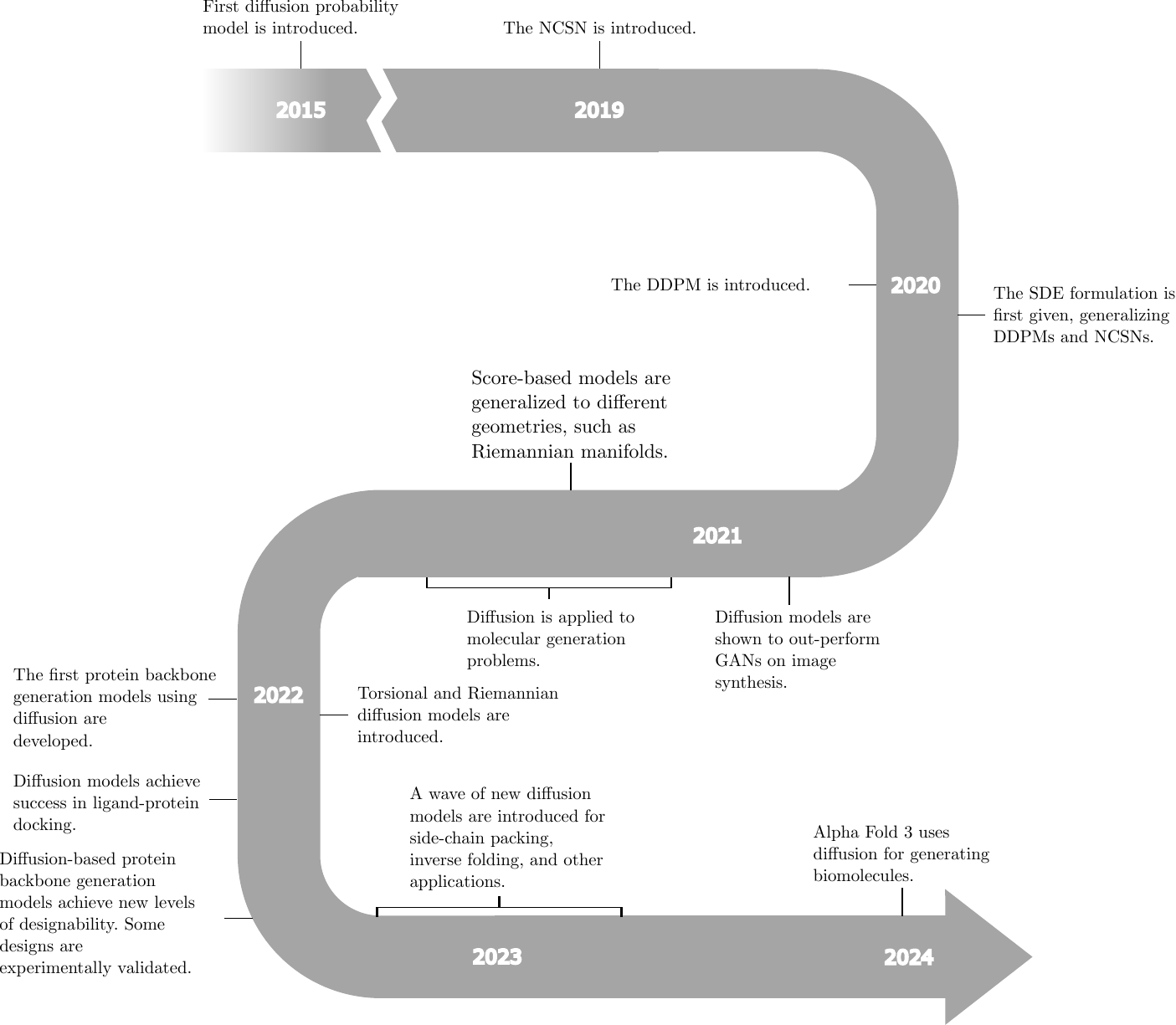}};
			\node[] at (5.3,13.97) {\footnotesize\cite{sohl-dickstein_deep_2015}};
			\node[] at (10.05,13.97) {\footnotesize\cite{song_generative_2019}};
		    \node[] at (11.6,10.36) {\footnotesize\cite{ho_denoising_2020}};
			\node[] at (16.44,9.65) {\footnotesize\cite{song_score-based_2020}};
			\node[] at (11.27, 4.8) {\footnotesize\cite{dhariwal_diffusion_2021}};
			\node[] at (8.1,5.1) {\footnotesize\cite{xu2022geodiff,shi2021learning}};
			\node[] at (9.65,8.03) {\footnotesize\cite{de_bortoli_riemannian_2022}};
			\node[] at (6.73,4.03) {\footnotesize\cite{jing_torsional_2023,huang2022riemannian}};
			\node[] at (1.8,3.99) {\footnotesize\cite{anand_protein_2022,trippe_diffusion_2023}};
			\node[] at (1.59,2.79) {\footnotesize\cite{corso_diffdock_2023,qiao_state-specific_2023}};
			\node[] at (2.95,0.79) {\footnotesize\cite{watson_novo_2023}};
			\node[] at (13.1,2.10) {\footnotesize\cite{abramson2024accurate}};
		\end{tikzpicture}
		\caption{Timeline of applications of diffusion models to biomolecules.}
	\end{figure}
}{
		\begin{figure}[h]
		\begin{tikzpicture}
			\node[anchor=south west,inner sep=0] at (0,0) {\includegraphics[width=\textwidth]{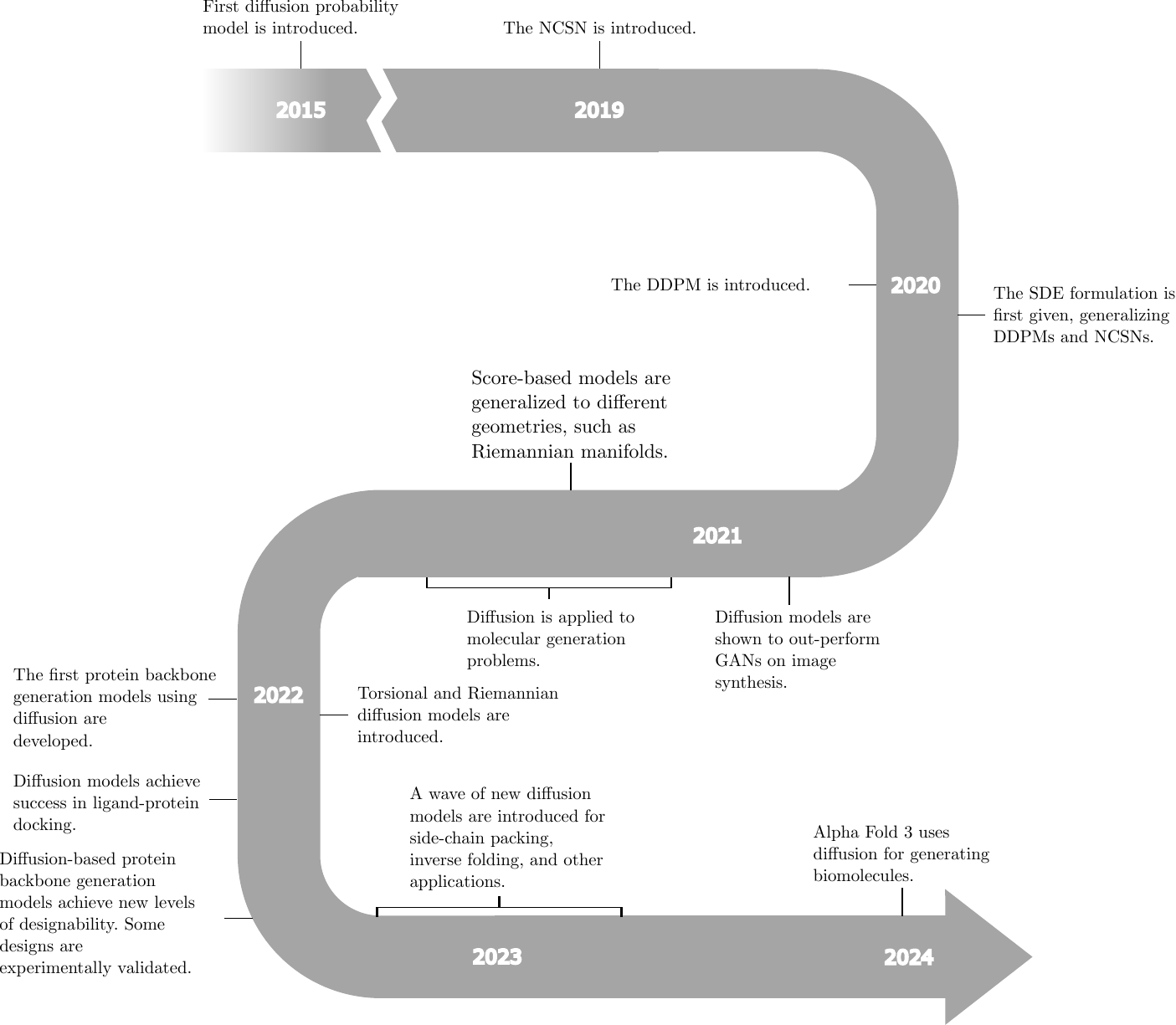}};
			\node[] at (5.3,13.97) {\footnotesize\cite{sohl-dickstein_deep_2015}};
			\node[] at (10.05,13.97) {\footnotesize\cite{song_generative_2019}};
			\node[] at (11.6,10.36) {\footnotesize\cite{ho_denoising_2020}};
			\node[] at (16.44,9.65) {\footnotesize\cite{song_score-based_2020}};
			\node[] at (11.27, 4.8) {\footnotesize\cite{dhariwal_diffusion_2021}};
			\node[] at (8.1,5.1) {\footnotesize\cite{xu2022geodiff,shi2021learning}};
			\node[] at (9.65,8.03) {\footnotesize\cite{de_bortoli_riemannian_2022}};
			\node[] at (6.73,4.03) {\footnotesize\cite{jing_torsional_2023,huang2022riemannian}};
			\node[] at (1.8,3.99) {\footnotesize\cite{anand_protein_2022,trippe_diffusion_2023}};
			\node[] at (1.59,2.79) {\footnotesize\cite{corso_diffdock_2023,qiao_state-specific_2023}};
			\node[] at (2.95,0.79) {\footnotesize\cite{watson_novo_2023}};
			\node[] at (13.1,2.10) {\footnotesize\cite{abramson2024accurate}};
		\end{tikzpicture}
		\caption{Timeline of applications of diffusion models to biomolecules.}
	\end{figure}
}

%% file: sections/motifs.tex
\section{Motifs for diffusion applied to biomolecules}

Due to the geometry of biomolecules and the common goals of researchers, when diffusion is applied to problems in the field several techniques and ideas come up often. Some of these are standard notions for diffusion generally, but others are more niche. This section covers ideas that appear frequently in research with the aim of highlighting the most important concepts when engaging in the literature.

\subsection{Controllable generation}

Unconditionally sampling from the learned distribution of a diffusion model is usually insufficient for applications. For example, in protein design, one may desire not just a protein sequence and conformation but also that the final structure satisfies some prescribed function. The SDE formulation of diffusion models allows for a straight-forward description of controllable generation. Given an observation \(\mathbf y\), the goal is to sample from the distribution of \(\mathbf x(0)\) conditioned on \(\mathbf y\), i.e., the distribution \(p_0(\mathbf x(0) \mid \mathbf y)\). The score of this distribution when perturbed can be rewritten using Bayes' rule to get 
\begin{equation*}
	\nabla_{\mathbf x} \log p_t (\mathbf x (t) \mid \mathbf y) = \nabla_{\mathbf x} \log p_t(\mathbf x(t)) + \nabla_{\mathbf x} \log p(\mathbf y \mid \mathbf x(t)).
\end{equation*}
Typically, \(\nabla_{\mathbf x} \log p_t(\mathbf x(t))\) will be approximated by a score network and so \(\nabla_{\mathbf x} \log p(\mathbf y \mid \mathbf x(t))\) will need to be estimated.

One option is to train another network to approximate this term. Training of the networks can often be done simultaneously so that the conditioned and unconditioned scores can be learned together. In some cases, domain knowledge or heuristics can be used to get an approximation. In the particular case of in-painting, the \emph{replacement method} has been used to approximate the conditional gradient. However, as pointed out in \cite{trippe_diffusion_2023}, this leads to an irreducible error, but a similar method using particle filtering can guarantee an approximation with the correct limiting distribution.

\subsection{Equivariant/Invariant score networks}

For many problems involving biomolecules, the space is three-dimensional Euclidean space (that is, \(\mathbb R^3\)) and the results should not depend on the choice of coordinate axes. For example, when predicting the native structure of a protein, two conformations are equivalent if one can be transformed into the other by translations and rotations. To be precise, the solution should be \emph{equivariant} (or \emph{invariant}) to rigid motions. A function \(f : \mathcal X \to \mathcal Y\) is equivariant with respect to a group \(G\) if for each \(g\in G\) and \(x \in \mathcal X\)
\begin{equation*}
	f(g\cdot x) = g \cdot f(x),
\end{equation*}
where \(g\cdot x\) and \(g\cdot f(x)\) represents a group action on \(\mathcal X\) and \(\mathcal Y\), respectively. A function is invariant if \(f(g\cdot x) = f(x)\) for each \(g \in G\) and \(x \in \mathcal X\). Restricting machine learning architectures to the smaller space of equivariant functions speeds up training and makes models robust to such transformations. Thus, networks that are equivariant with respect to the special Euclidean group \(\se3\) are a natural choice for problems involving biomolecules\footnote{Due to the chirality of certain biomolecules, one typically does not want equivariance to \emph{reflections}. Hence, equivariance/invariance is typically with respect to \(\se 3\) and not \(\mathrm{E}(3)\)}.

A common choice for incorporating equivariance into diffusion models is to use an equivariant network for \(s_\theta(\mathbf x, t)\). Given that the diffusion process and the score are both equivariant with respect ot \(\se3\), the reverse diffusion and sampling processes are also equivariant (c.f.\ \cite[Prop.~1]{xu2022geodiff} for instance). Common choices of architecture for the score network include Invariant Point Attention \cite{jumper_highly_2021}, Tensor Field Networks \cite{thomas2018tensor}, and Equivariant Graph Neural Networks \cite{satorras_en_2021}, among other.

\subsection{Diffusion on manifolds}

While equivariance provides numerous benefits when designing diffusion models, it may also lead unrealistic generation of molecules. This is due to the chirality of some biomolecules, which is not preserved under reflections of \(\mathbb R^3\). Furthermore, for molecules with \(n\) particles, the effective number of degrees of freedom is typically much smaller than \(3n\). For instance, bond lengths and angles are usually fairly rigid in molecules and thus contribute little to the diversity of conformations. Thus, one can imagine the configurations of a biomolecule as lying on a smaller submanifold of the ambient space, and by defining coordinates on that manifold the configuration of the molecules can be specified. It is possible to define the diffusion process on these new coordinates, which reduces the dimension of the problem while enforcing certain geometric priors.

Score-based diffusion models may be carried over to Riemannian manifolds while retaining many of the characteristics of the models from Cartesian space \cite{de_bortoli_riemannian_2022}. A nautral choice for forward diffusion on compact manifolds is the VE SDE
\begin{equation*}
	\mathrm d \mathbf x = \sqrt{\frac{\mathrm d [\sigma(t)^2]}{\mathrm d t}}\mathrm d \mathbf w_{\mathcal{M}},
\end{equation*} 
where \(\mathbf w_{\mathcal M}\) is now Brownian motion on the manifold. This SDE has a uniform distribution as its stationary distribution. For generic Riemannian manifolds, the transition kernel cannot be explicitly written, and so training may be computationally expensive. However, the kernel is known for some simple manifolds. For rigid bodies in three dimensions, two manifolds are particularly useful: the torus \(\mathbb T\), and the set of orientations \(\so3\).

\begin{figure}[ht]
	\centering
	\includegraphics[scale=0.65]{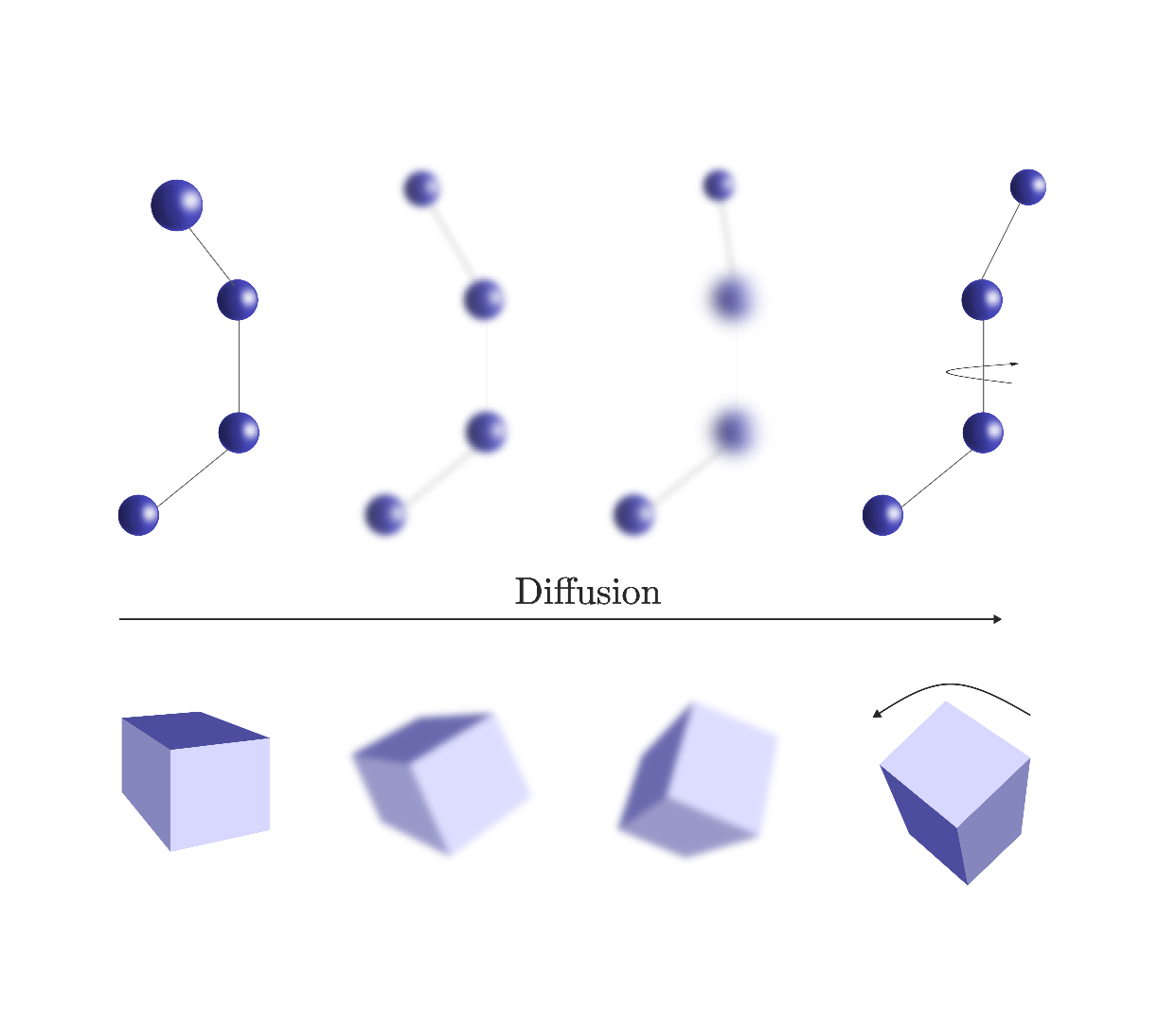}
	\vspace{-10pt}
	\caption{Sample paths for diffusion in \(\mathbb T\) and \(\so3\). (top) Diffusion in \(\mathbb T\) is frequently applied to the torsion angle of four consecutive particles. Since bond lengths and angles are usually fairly rigid, most of the diversity in conformations can be explained by the torsional angles. Here an example torsional angle is perturbed by the diffusion process. (bottom) Diffusion in \(\so3\) can be applied to frame data to perturb the orientation. Here the cube's center is not moved while a rigid rotation is applied.}
\end{figure}

\subsubsection{Diffusion on \texorpdfstring{$\mathbb T$}{T}: torsional space}
%\subsubsection{Diffusion on $\mathbb T$: torsional space}

The manifold \(\mathbb T\) can be used to represent angular data, with the points being associated with the interval \([0,2\pi)\) \cite{jing_torsional_2023}. The conformation for many molecules can almost entirely be described by their torsional angles, and so modeling the diffusion process on \(\mathbb T^m\) can effectively reduce the dimensionality without a significant loss of accuracy.

For the VE formulation the transition kernel on \(\mathbb T^m\) is given by the wrapped Gaussian:
\begin{equation*}
	p_t(\bm\tau \mid \bm\tau ') \propto \sum_{\mathbf k\in \mathbb Z^m} \exp \left( \frac{-|\bm\tau - \bm \tau' - 2\pi \mathbf k|^2}{2\sigma(t)^2} \right)
\end{equation*}
Sampling from this distribution amounts from sampling from a normal distribution in \(\mathbb R^m\) and then modding out by units of \(2\pi\).

\subsubsection{Diffusion on \texorpdfstring{$\so3$}{SO(3)}: orientations in three dimensions}
%\subsubsection{Diffusion on $\so3$: orientations in three dimensions}
The manifold \(\so3\) is the set of \(3\times 3\) orthogonal matrices with determinant \(1\), and so they are used to represent rigid rotations of \(\mathbb R^3\) or the orientation of three-dimensional objects. Diffusing on \(\so3\) can be useful during generation when the structure being diffused is not spherical and properties are dependent on its orientation -- for example a ligand or the residue of a protein. The transition kernel for the VE formulation is given by the \(\igso\) distribution \cite{nikolayev_normal_1997}, which easily computable via a series expansion.

\subsubsection{Direct product of manifolds}

More complicated manifolds can be built from the direct product of simpler manifolds. The VE forward process diffuses each coordinate independently, and so for direct product the diffusion can be carried out separately on each component manifold. This allows for diffusion across multiple coordinates: for instance, diffusing the position and orientation of \(n\) particles amounts to diffusing on the manifold \(\mathbb R^{3n} \times \so3^n\), for which we know each of the transition kernels.

\subsection{Low-temperature sampling}

A common problem with training generative models is \emph{overdispersion}. When this occurs, the coverage of the modes of an underlying distribution is emphasized more than sample quality, leading to poor quality in the samples. This can be particularly deleterious when using generative models for predictions. It is common to use modified sampling to sacrifice diversity for higher-quality. See for instance shruken encodings in normalizing flows \cite{kingma_glow_2018}.

Low-temperature sampling of a distribution is an attractive way to combat overdispersion. That is, if \(p(x)\) is the original distribution, sampling instead from \(\frac 1 Z p(x)^\lambda\), where \(Z\) is a normalizing factor and \(\lambda > 0\) is an inverse temperature parameter, will emphasize high likelihood areas of the distribution. Taking \(\lambda \to \infty\) will lead the distribution to concentrate around the global maximum of \(p(x)\). This process is analogous to lowering the temperature for a thermodynamic system so that enthalpy is emphasized over entropy. However, low-temperature sampling in many cases is intractable or requires expensive Markov chain Monte Carlo (MCMC) simulations to compute.

For diffusion models, it is tempting to simply increase the score and decrease the noise, but doing this does not give the correct distribution and is generally ineffective \cite{dhariwal_diffusion_2021}. Ingraham et al.\ demonstrate how to derive an approximate scheme in the case of a normal distribution \cite{ingraham_illuminating_2023}. One can then approximate the perturbed score function \(\mathbf s_{\mathrm{perturb}}\) by multiplying the original score function \(\mathbf s\) by a time-dependent constant \(\lambda_t\): that is,
\begin{equation*}
	\mathbf s_{\mathrm{perturb}}(\mathbf x, t) \approx \lambda_t \mathbf s (\mathbf x, t). 
\end{equation*}
Replacing \(\mathbf s\) with \(\mathbf s_{\mathrm{perturb}}\) leads to new reverse diffusion SDE that one can use to sample the low-temperature distribution. This approximating only holds in the case of a Gaussian distribution for \(p_{\mathrm{data}}\), and so applying the rescaling to arbitrary distribution will not lead to a proper reweighting. However, mixing the low-temperature reverse SDE with a Langevin SDE allows for proper sampling of a generic distribution. For more details, we refer readers to \cite[Appendix C]{ingraham_illuminating_2023}.

%% file: sections/applications.tex
\section{Applications}
\ifthenelse{\boolean{isJMB}}{
	\begin{figure}[p]
		\begin{tikzpicture}
			\node[anchor=south west,inner sep=0] at (0,0) {\includegraphics[width=\textwidth]{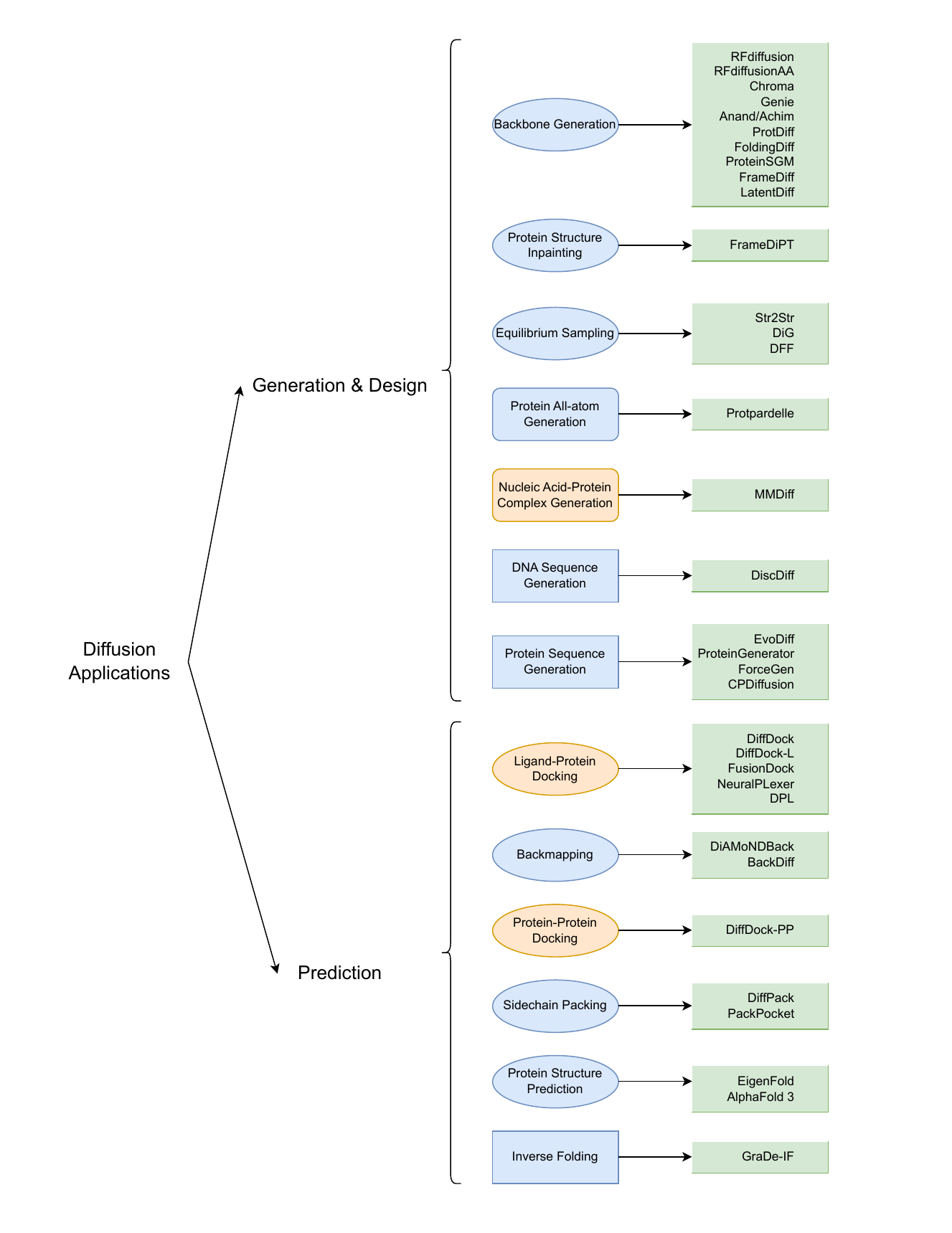}};
			\node[] at (14.05,20.335) {\tiny\cite{watson_novo_2023}};
			\node[] at (14.05,20.075) {\tiny\cite{krishna2024generalized}};
			\node[] at (14.05,19.815) {\tiny\cite{ingraham_illuminating_2023}};
			\node[] at (14.05,19.563) {\tiny\cite{lin_generating_2023}};
			\node[] at (14.05,19.302) {\tiny\cite{anand_protein_2022}};
			\node[] at (14.05,19.041) {\tiny\cite{trippe_diffusion_2023}};
			\node[] at (14.05,18.780) {\tiny\cite{wu_protein_2024}};
			\node[] at (14.05,18.519) {\tiny\cite{lee_score-based_2023}};
			\node[] at (14.05, 18.258) {\tiny\cite{yim_se3_2023}};
			\node[] at (14.05,17.995) {\tiny\cite{fu_latent_2023}};
			\node[] at (14.05,17.08) {\tiny\cite{zhang_framedipt_2024}};
			\node[] at (14.05,15.84) {\tiny\cite{lu_str2str_2023}};
			\node[] at (14.05,15.579) {\tiny\cite{zheng2024predicting}};
			\node[] at (14.05,15.295) {\tiny\cite{arts2023two}};
			\node[] at (14.05,14.17) {\tiny\cite{chu_all-atom_2023}};
			\node[] at (14.05,12.77) {\tiny\cite{morehead_towards_2023}};
			\node[] at (14.05,11.38) {\tiny\cite{li_latent_2023}};
			\node[] at (14.05,10.285) {\tiny\cite{alamdari_protein_2023}};
			\node[] at (14.05,10.024) {\tiny\cite{lisanza_joint_2023}};
			\node[] at (14.05,9.755) {\tiny\cite{ni_forcegen_2023}};
			\node[] at (14.05,9.485) {\tiny\cite{zhou_conditional_2023}};
			\node[] at (14.05,8.555) {\tiny\cite{corso_diffdock_2023}};
			\node[] at (14.05,8.294) {\tiny\cite{corso2024deep}};
			\node[] at (14.05,8.033) {\tiny\cite{masters_fusiondock_nodate}};
			\node[] at (14.05,7.772) {\tiny\cite{qiao_state-specific_2023}};
			\node[] at (14.05,7.511) {\tiny\cite{nakata_end--end_2023}};
			\node[] at (14.05,6.685) {\tiny\cite{jones2023diamondback}};
			\node[] at (14.05,6.405) {\tiny\cite{liu2023backdiff}};
			\node[] at (14.05,5.235) {\tiny\cite{ketata_diffdock-pp_2023}};
			\node[] at (14.05,4.065) {\tiny\cite{zhang_diffpack_2023}};
			\node[] at (14.05,3.765) {\tiny\cite{zhang2024packdock}};
			\node[] at (14.05,2.615) {\tiny\cite{jing_eigenfold_2023}};
			\node[] at (14.05,2.335) {\tiny\cite{abramson2024accurate}};
			\node[] at (14.05,1.33) {\tiny\cite{yi_graph_2023-1}};
		\end{tikzpicture}
		\caption{Summary of diffusion application for the generation and prediction of biomolecules. Ellipses are used for structure generation/prediction; rectangles for sequences; and rounded rectangles for co-generation methods. Blue shapes represent monomeric applications while orange shapes represent polymeric/complex applications.}
	\end{figure}
}{
		\begin{figure}[p]
		\begin{tikzpicture}
			\node[anchor=south west,inner sep=0] at (0,0) {\includegraphics[width=\textwidth]{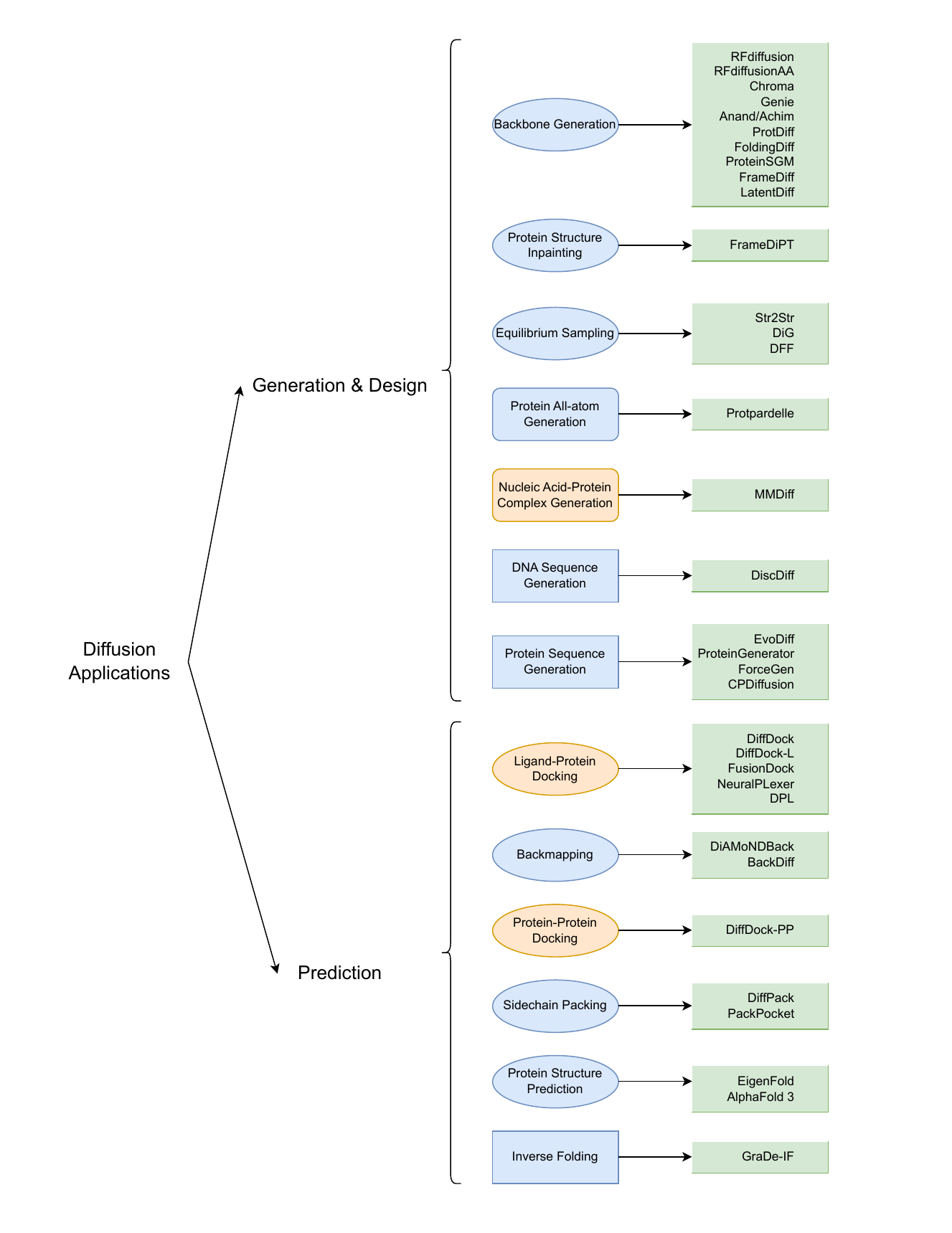}};
			\node[] at (14.05,20.36) {\tiny\cite{watson_novo_2023}};
			\node[] at (14.05,20.10) {\tiny\cite{krishna2024generalized}};
			\node[] at (14.05,19.84) {\tiny\cite{ingraham_illuminating_2023}};
			\node[] at (14.05,19.588) {\tiny\cite{lin_generating_2023}};
			\node[] at (14.05,19.327) {\tiny\cite{anand_protein_2022}};
			\node[] at (14.05,19.066) {\tiny\cite{trippe_diffusion_2023}};
			\node[] at (14.05,18.805) {\tiny\cite{wu_protein_2024}};
			\node[] at (14.05,18.544) {\tiny\cite{lee_score-based_2023}};
			\node[] at (14.05, 18.283) {\tiny\cite{yim_se3_2023}};
			\node[] at (14.05,18.02) {\tiny\cite{fu_latent_2023}};
			\node[] at (14.05,17.11) {\tiny\cite{zhang_framedipt_2024}};
			\node[] at (14.05,15.84) {\tiny\cite{lu_str2str_2023}};
			\node[] at (14.05,15.579) {\tiny\cite{zheng2024predicting}};
			\node[] at (14.05,15.295) {\tiny\cite{arts2023two}};
			\node[] at (14.05,14.19) {\tiny\cite{chu_all-atom_2023}};
			\node[] at (14.05,12.77) {\tiny\cite{morehead_towards_2023}};
			\node[] at (14.05,11.38) {\tiny\cite{li_latent_2023}};
			\node[] at (14.05,10.31) {\tiny\cite{alamdari_protein_2023}};
			\node[] at (14.05,10.049) {\tiny\cite{lisanza_joint_2023}};
			\node[] at (14.05,9.78) {\tiny\cite{ni_forcegen_2023}};
			\node[] at (14.05,9.51) {\tiny\cite{zhou_conditional_2023}};
			\node[] at (14.05,8.58) {\tiny\cite{corso_diffdock_2023}};
			\node[] at (14.05,8.319) {\tiny\cite{corso2024deep}};
			\node[] at (14.05,8.058) {\tiny\cite{masters_fusiondock_nodate}};
			\node[] at (14.05,7.797) {\tiny\cite{qiao_state-specific_2023}};
			\node[] at (14.05,7.536) {\tiny\cite{nakata_end--end_2023}};
			\node[] at (14.05,6.71) {\tiny\cite{jones2023diamondback}};
			\node[] at (14.05,6.43) {\tiny\cite{liu2023backdiff}};
			\node[] at (14.05,5.26) {\tiny\cite{ketata_diffdock-pp_2023}};
			\node[] at (14.05,4.09) {\tiny\cite{zhang_diffpack_2023}};
			\node[] at (14.05,3.79) {\tiny\cite{zhang2024packdock}};
			\node[] at (14.05,2.64) {\tiny\cite{jing_eigenfold_2023}};
			\node[] at (14.05,2.36) {\tiny\cite{abramson2024accurate}};
			\node[] at (14.05,1.33) {\tiny\cite{yi_graph_2023-1}};
		\end{tikzpicture}
		\caption{Summary of diffusion application for the generation and prediction of biomolecules. Ellipses are used for structure generation/prediction; rectangles for sequences; and rounded rectangles for co-generation methods. Blue shapes represent monomeric applications while orange shapes represent polymeric/complex applications.}
	\end{figure}
}

Applications for diffusion models can be roughly divided into two categories: generation/design and prediction. Methods between these two categories are similar, but the goals are different in terms of sampling the underlying distribution. For generative applications, the goal is to faithfully sample the distribution. This means not only having accurate samples but also having good coverage of all modes. Design tasks are usually downstream from generation task, where one conditionally samples the distribution of interest. For predictive applications, the goal is to sample the most likely point in the probability distribution. This tends to correspond with an optimization goal, e.g., minimizing the total energy of a system.

This classification is not a strict binary, but rather represent two ends of a spectrum for sampling a distribution. There is a tension between sample \emph{diversity} (how well a method covers all modes of a distribution) and sample \emph{quality} (the likelihood of samples from a distribution), and there may be cases where one might sacrifice sample diversity for better quality or vice versa. For instance, we discussed how low-temperature sampling gives a controlled way to ignore smaller modes in the distribution to instead focus on more likely points. Even in prediction cases, where a single answer is desired, diversity can still be beneficial. Diversity of sampling can benefit prediction when (a) there are multiple plausible solutions, (b) there are close modes which must be discriminated or (c) a measure of flexibility or certainty is needed.

\begin{figure}[h]
	\centering
	\includegraphics[width=\textwidth]{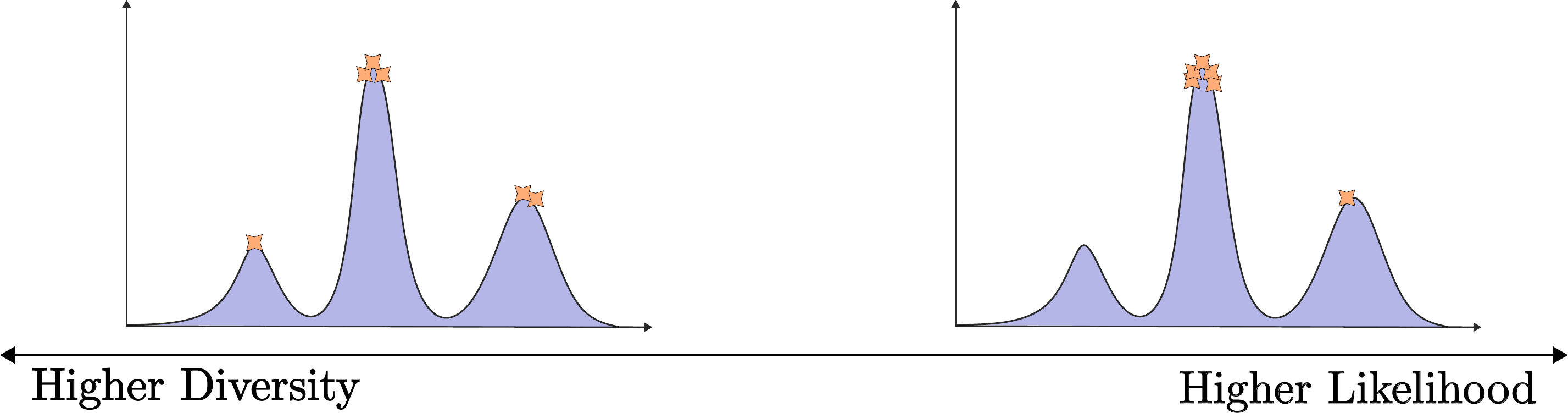}
	\caption{The difference between generation tasks and prediction tasks is how samples are to be drawn from the distribution. On the one hand, generation tasks value faithful sampling of the distribution and the sampling of all modes. On the other hand, prediction tasks value only the most likely outcomes from a distribution. The left- and right-hand plots above show this dichonomy between generation and prediction, respectively. Generated samples from the methods are shown in orange stars. }
\end{figure}

\subsection{Generative modeling}

Diffusion models are particularly useful for generating samples from a high-dimensional distribution, and so it is natural to apply them to the problem of creating plausible molecular  structures. Furthermore, conditional diffusion allows for additional constraints to be added to the generation process, allowing for the design of proteins. Despite the recentness of the techniques, there have already been significant results. Many results focus on generating the backbone of the protein sequence (for example the locations of the C\(\alpha\) atoms), and so this direction will be the main focus of this section.

\subsubsection{Benchmarking}

The ability of generative models to effectively sample the desired distribution is typically measured in three aspects: sample quality, sample diversity, and novelty. \emph{Sample quality} looks at whether the generated samples are likely to come from the underlying distribution. \emph{Sample diversity} looks at whether the distribution of samples reasonably spans the entirety of the underlying distribution. \emph{Novelty} is a measure of how well the model generalizes from the training set; a model that simply memorizes the training data would do well with quality and diversity, but would not be able to produce new samples from the distribution.

To measure the quality of generated protein structures, one needs to determine whether a polypeptide chain has a native conformation matching the generated structure. By design, the generated structures are novel and thus cannot be directly compared with previous structures. Validating these structures experimentally is time-intensive, and so \emph{in silico} methods for determining the structure -- such as AlphaFold2 \cite{jumper_highly_2021} and ESMFold \cite{lin2023evolutionary} -- are employed to evaluate the samples. \emph{Self-consistency} of a method is the measure of whether a sequence can be found and independently folded into the generated structure. For many models, no sequence is generated and so a reverse folding method, such as ProteinMPNN \cite{dauparas2022robust}, must first be used to find a number of plausible sequences. The self-consistency score is computed by comparing the folded structure with the generated structure and taking the closest comparison. The self-consistency TM (scTM) score computes the TM score between the generated backbone and each folded backbone and returns the highest result. Another choice is the self-consistency RMSD (scRMSD) score which similarly compares structures using the backbone RMSD and returns the smallest result. The scRMSD is a more stringent requirement, and is a stricter measurement of designability. A protein is said to be \emph{designable} if its scRMSD is less than 2 Å or its scTM score is greater than \(0.5\).

Assessing diversity and novelty is more straight-forward since this only involves comparisons with the generated and known structures. The average pairwise TM score among the generated structures measures the diversity of the sampling, and the maximum TM score of the generated structure and the training set measures the novelty of the sampling. Other qualitative measures of assessing the model exist as well. For unconditional generation, comparing the distribution of relevant statistics (e.g.\ dihedral angles and secondary structure) can help determine whether the sampling does not well approximate the underlying distribution.

\subsubsection{Backbone generation models}

Trippe et al.\ created a generative model for the C\(\alpha\) backbone of proteins with a specific focus on controllable generation for the motif-scaffold problem \cite{trippe_diffusion_2023}. Diffusion was carried out on the Cartesian coordinates and used an \(E(3)\) equivariant score network. Despite producing plausible protein structures, generally the backbones were not designable. Furthermore, the equivariance of the network to reflections produced incorrect left-handed \(\alpha\)-helices, and so the model did not respect the chirality of the proteins. Wu et al.\ diffused on the internal angles of the protein backbone in order to create a generative model that respects chirality \cite{wu_protein_2024}. Again, the generated proteins were plausible and had distributions of secondary structures similar to native proteins, but most generated backbones were not designable. 

ProteinSGM was one of the first models to be able to generate highly designable protein backbones \cite{lee_score-based_2023}. The diffusion is carried out on the six-dimensional space of inter-residue coordinates as defined in trRossetta \cite{yang2020improved}, using the VE SDE formulation. Protein backbones are obtained from coordinates using an adaptation of the trRossetta minimization method, which enables generation of realistic structures. ProteinSGM is able to generate proteins up to length 128 with scTM of 90.2\%, versus 11.8\% and 22.7\% as reported by \cite{trippe_diffusion_2023} and \cite{wu_protein_2024}, respectively.

Ingraham et al.\ developed Chroma, which generates backbones by diffusing on the C\(\alpha\) atoms \cite{ingraham_illuminating_2023}. The backbones generated are more likely to be designable than the previously mentioned methods, with around 50\% being designable by their criterion. Notable features of their model include (1) choosing a covariance structure of the initial distribution to replicate the statistics of realistic protein backbones (such as radius of gyration), (2) a random graph neural score network to maintain subquadratic computational time, and (3) an annealed sampling scheme that allows the model to sample from \(\frac 1 Z p(x)^\lambda\) without retraining. Furthermore, the programmability of the model and code is emphasized in the paper. That is, several conditioners are created to allow for a wide range of conditions for protein generation, such as on the three-dimensional shape and point symmetries of the final structure. The generated protein structures are diverse and cover all of natural protein space.

Other recent attempts at diffusion models for protein generation carry out the diffusion process on the coordinates and orientations of the backbone atoms. That is, for \(n\) atoms, diffusion will be carried out on \(\mathbb{R}^{3n} \times \so3^{n} \simeq \se3\). This allows the models to reason about the three-dimensional structure of the protein during the generation step while retaining angular information between residues so that chirality is respected. This paired with \(\se3\) equivariant networks like IPA have been particularly successful.

The most salient example of this method is RFdiffusion, which is considered the gold standard of backbone generation \cite{watson_novo_2023}. As argued by the authors, previous models lacked deep understanding of protein structure, and so taking advantage of structure predictions methods could allow for better sample quality. In particular, pretraining using the weights from RoseTTAFold \cite{baek2021accurate} provided an enormous benefit to the diffusion model. RFdiffusion has high designabiltiy of unconditionally generated proteins, with independent benchmarking showing over \(95\%\) of moderately sized backbones being designable \cite{bose2023se,lin_generating_2023} and some of the protein designs being experimentally verified \cite{watson_novo_2023}. Furthermore, the model is capable of conditional generation of proteins, such as producing oligomers with point symmetries and functional-motif scaffolding. Recent work has been done on expanding RFdiffusion's capabilities to more general biomolecular tasks. RoseTTAFold All-atom (RFAA) \cite{krishna2024generalized} expands RoseTTAFold's by including atomic-level inputs for other biomolecular components. This allows RFAA to more accurately predict protein structure in the context of assemblies of molecules. RFdiffusionAA (similar to the original RFdiffusion) uses the pretrained weights of RFAA to develop a protein backbone generation model. Thus RFdiffusionAA can generate protein structures conditioned on small-molecule binders. Designs from RFdiffusionAA were experimentally validated and were shown to bind with the target molecules.

\ifthenelse{\boolean{isJMB}}{
	\begin{figure}[h]
	\centering
	\begin{subfigure}[t]{0.45\textwidth}
		\centering
		\includegraphics[width=\linewidth]{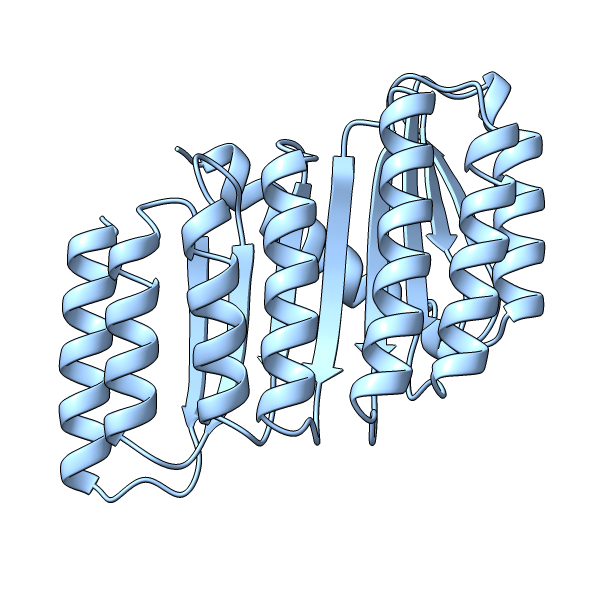}
		\caption{} \label{fig:rfdiffusion}
	\end{subfigure}
	\hfill
	\begin{subfigure}[t]{0.45\textwidth}
		\centering
		\includegraphics[width=\linewidth]{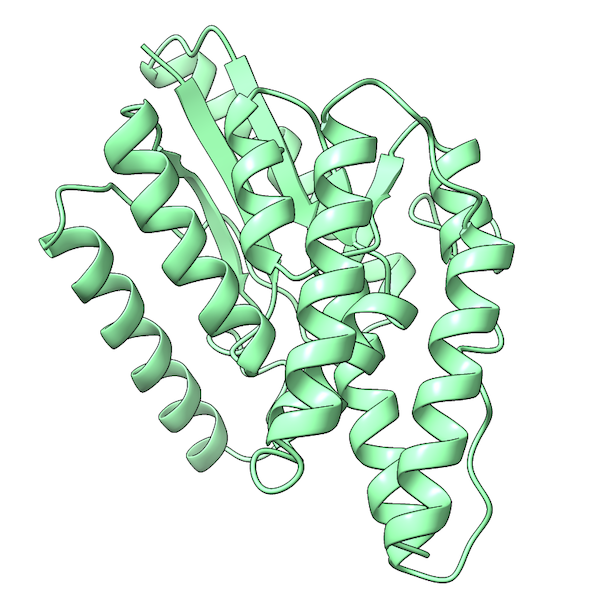}
		\caption{} \label{fig:chroma}
	\end{subfigure}
	\vspace{1cm}
	\begin{subfigure}[t]{0.45\textwidth}
		\centering
		\includegraphics[width=\linewidth]{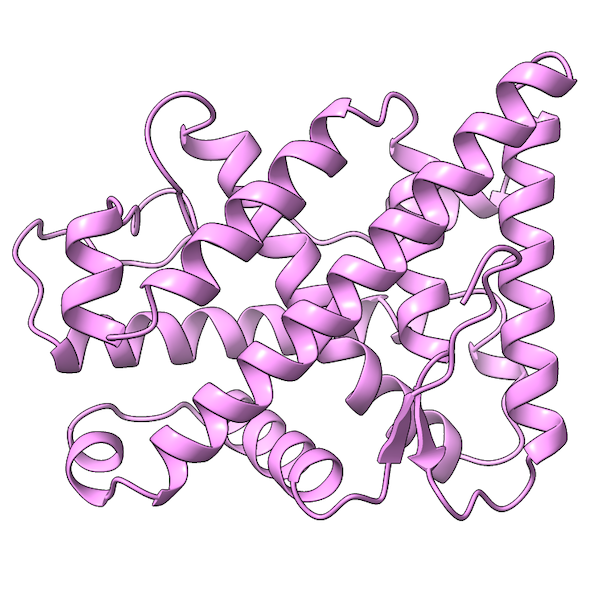}
		\caption{} \label{fig:framediff}
	\end{subfigure}
	\hfill
	\begin{subfigure}[t]{0.45\textwidth}
		\centering
		\includegraphics[width=\linewidth]{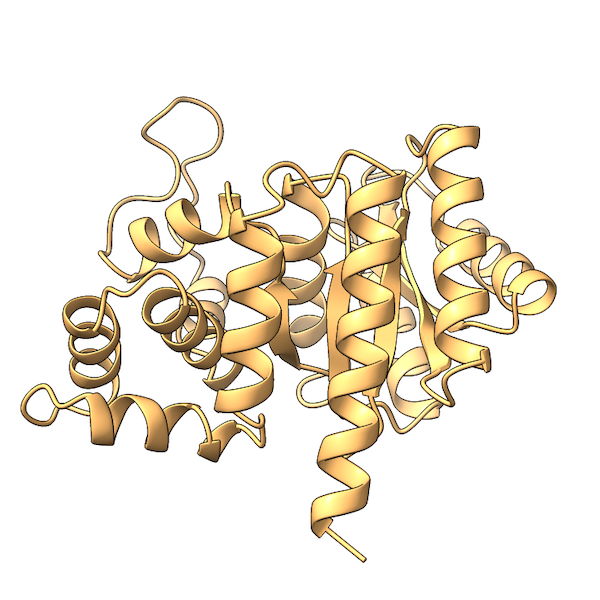}
		\caption{} \label{fig:genie}
	\end{subfigure}
	\caption{Example protein backbone generated from diffusion methods. Each backbone is 250 residues long:  (a) RFdiffusion \cite{watson_novo_2023}; (b) Chroma \cite{ingraham_illuminating_2023}; (c) FrameDiff \cite{yim_se3_2023}; (d) Genie \cite{lin_generating_2023}.}
\end{figure}}
{
		\begin{figure}[p]
		\centering
		\begin{subfigure}[t]{0.45\textwidth}
			\centering
			\includegraphics[width=\linewidth]{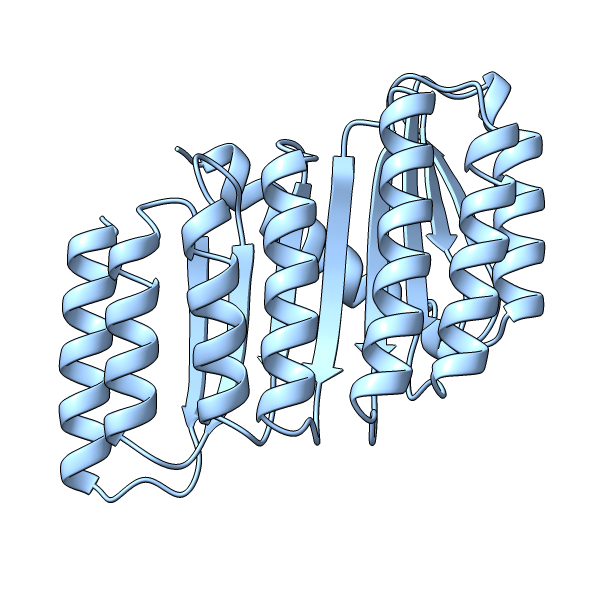}
			\caption{} \label{fig:rfdiffusion}
		\end{subfigure}
		\hfill
		\begin{subfigure}[t]{0.45\textwidth}
			\centering
			\includegraphics[width=\linewidth]{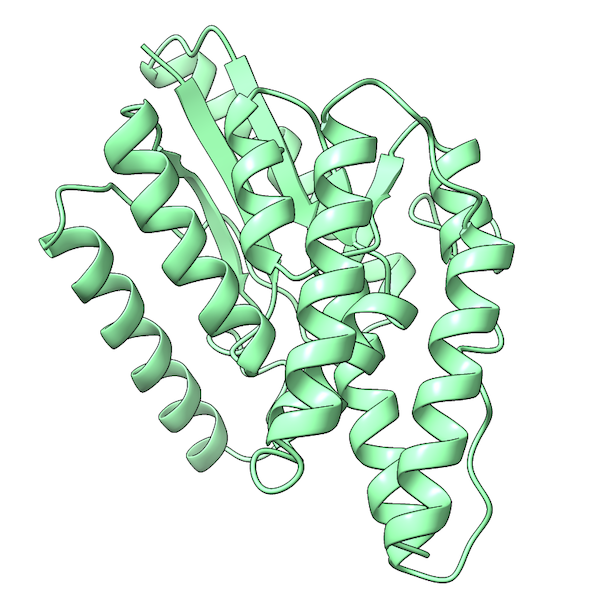}
			\caption{} \label{fig:chroma}
		\end{subfigure}
		\vspace{1cm}
		\begin{subfigure}[t]{0.45\textwidth}
			\centering
			\includegraphics[width=\linewidth]{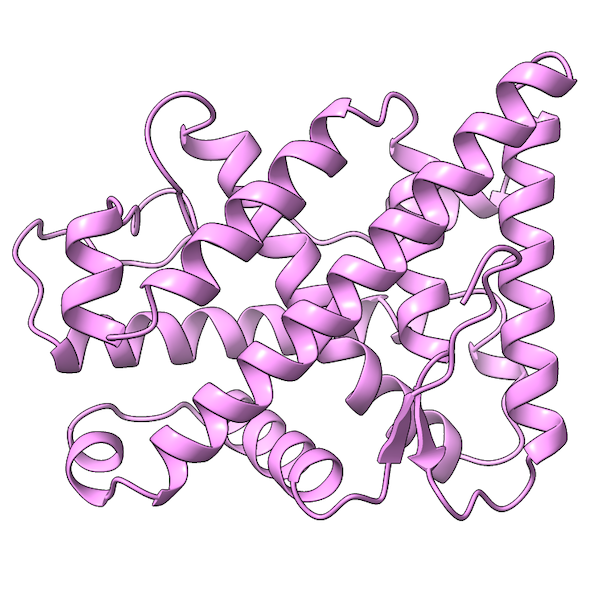}
			\caption{} \label{fig:framediff}
		\end{subfigure}
		\hfill
		\begin{subfigure}[t]{0.45\textwidth}
			\centering
			\includegraphics[width=\linewidth]{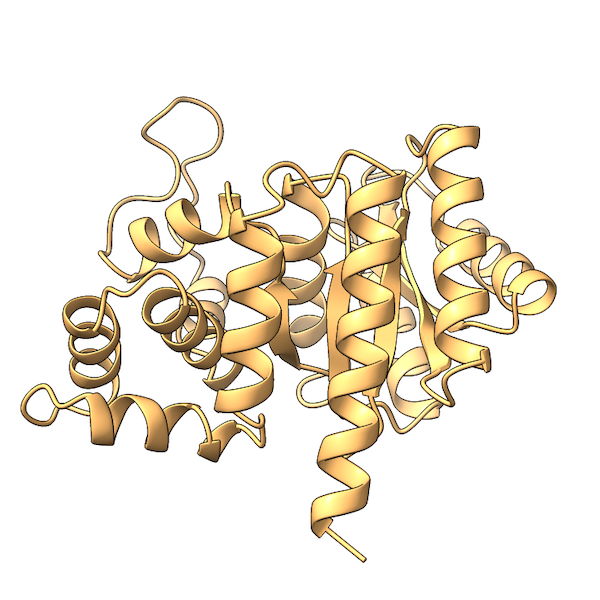}
			\caption{} \label{fig:genie}
		\end{subfigure}
		\caption{Example protein backbone generated from diffusion methods. Each backbone is 250 residues long:  (a) RFdiffusion \cite{watson_novo_2023}; (b) Chroma \cite{ingraham_illuminating_2023}; (c) FrameDiff \cite{yim_se3_2023}; (d) Genie \cite{lin_generating_2023}.}
		\end{figure}
}

Other models have used \(\se3\) diffusion to get designable backbones without the need to pretrain. Anaand and Achim were able to generate plausible protein structures diffusing on \(\se3\)\footnote{More accurately the diffusion took place on \(\mathbb R^3 \times \mathrm{SU}(2)\), where quaternions are used to represent orientation.} and capture realistic distributions of secondary structures \cite{anand_protein_2022}. The diffusive process in this case was approximated by random interpolations in the space. FrameDiff uses a principled approach of diffusing on \(\se3\) where the loss is given by denoising score matching (as opposed to a Frobenius norm used in \cite{watson_novo_2023}). The score network uses IPA to capture spatial relationships while a transformer captures interactions along the protein chain. The model is able to generate designable backbones (using the scTM metric) about \(75\%\) of the time, while still maintaining diversity and novelty of the samples. This model is also extendable to the protein in-painting problem, as demonstrated by FrameDiPT \cite{zhang_framedipt_2024}.

Lin and AlQuraishi carry out a similar idea by using \(\se3\) equivariant network \cite{lin_generating_2023}. The forward diffusion process is carried out on the coordinates on the C\(\alpha\) atoms while the backward process uses coordinate frames to predict the noise displacements at each step. This allows for a cheaper noising process while allowing the full reasoning of IPA when generating samples. The authors demonstrate improved designability and diversity of tertiary structures compared with FrameDiff (although FrameDiff showed better diversity of secondary structure elements).

\begin{table}[ht]
	\centering
	\begin{adjustbox}{max width=\linewidth}
		\begin{tabular}{lc|c|c|c}
			\toprule
			& Designability & Diversity & Novelty  & \\
			Method  & \% scRMSD $<$ 2.0 Å $\uparrow$ & TM-Score $\downarrow$ & \% PDBTM $<$ 0.5 $\uparrow$ & Number of Parameters \\
			\midrule
			RFdiffusion & \textbf{0.904} & 0.2894 & 0.3750  & 59.8 M\\
			Chroma & 0.728 &  0.2779 & 0.3198 & 32.4 M \\
			FrameDiff & 0.484 & 0.2912 &  0.157 & 17.4 M\\
			Genie & 0.668 &\textbf{0.2674} & \textbf{0.5069} & 4.1 M\\
			\bottomrule
		\end{tabular}
	\end{adjustbox}
	\vspace{3pt}
	\caption{Benchmarking of backbone generation models on unconditional generation.}
	\label{tab:backbone}
\end{table}

For convenience, we provided original benchmarking of backbone generation methods. The results are given  in \cref{tab:backbone} and further details of the benchmarking are provided in \cref{app:backbone}. Notably, RFdiffusion performs the best in designability, which matches the benchmarking done in other works. Genie performs well with diversity and novelty, despite its relatively small size compared to the other models. One should note that there is a tradeoff between designability and diversity/novelty: prioritizing designability will lead to proteins with more stable structures but may bias toward predictable designs (e.g.\ an overemphasis on \(\alpha\)-helices). However, RFdiffusion still performs well on all metrics, even given its high designability. The strength of RFdiffusion can likely be credited to using pretrained weights from RoseTTAFold, which gives useful information about protein tertiary structure.

\subsubsection{Other generative models}

Although generating proteins based on backbones coordinates is a popular choice, other approaches have been attempted.  One deficit to focusing on coordinates is that conditional generation based on sequence information is not possible: the backbone methods above only consider structural information and obtain the sequences via a reverse folding method. Thus diffusing in \emph{sequence} space could be useful for conditioning on sequence information. Up to this point, we have only discussed diffusion on continuous spaces (either Euclidean space or a manifold), while diffusing on sequences requires being in a discrete space. One option for bringing diffusion to discrete space is to simply embed categorical information into continuous (e.g.\ using a one-hot encoding scheme) and then performing a traditional diffusion. A more principled options is to create a diffusion scheme for discrete data. For instance, Discrete Denoising Diffusion Probabilistic Models (D3PMs) describe a corruption process for discrete data that is analogous to Gaussian noise \cite{austin_structured_2023}.

This more principled approach was taken by the authors of EvoDiff, which is a protein sequence generation model \cite{alamdari_protein_2023}. EvoDiff carries out discrete diffusion in a method similar to D3PM, but the authors found it more successful to generate the sequence autoregressively. The model demonstrates good coverage of protein sequence space. By embedding the sequences with ProtT5 \cite{elnaggar_prottrans_2022} and computing the Fr\'echet distance between the generated set and the test set, the authors show that EvoDiff provides better coverage than other protein language models \cite{verkuil_language_2022} and RFdiffusion \cite{watson_novo_2023}. EvoDiff is able to generate proteins with intrinsically disordered regions, which is something that other structure-based diffusion methods struggle to do. Furthermore, it is also possible to condition on MSA data to generate other members of a protein family without retraining. Overall, many of the strengths of EvoDiff are orthogonal to those of structure-based diffusion.

Protein Generator is another method that opts to diffuse over sequence space to generate protein structures \cite{lisanza_joint_2023}. However, the diffusion process is carried out in continuous space as standard DDPM, where the amino acid sequence is embedded via a one-hot encoding. Structures are simultaneously generated by fine-tuning RoseTTAfold \cite{baek2021accurate} to accept noisy sequences and diffusion times\footnote{This method of structure generation echoes RFdiffusion, which uses pretrained weights from RoseTTAfold to guide training.}. This method creates sequences alongside a physically plausible three-dimensional structure. This also allows for the conditional generation based on sequence and structure.

Research has also been focused on sequence generation of proteins with predefined functions. For instance, ForceGen uses a pretrained protein language (pLM) model to diffuse in probability space \cite{ni_forcegen_2023}. The authors were particularly focused on generating proteins which matched mechanical folding responses. To accomplish this, they curated a dataset consisting of protein sequences as well MD simulation data of the force required to stretch a protein. Samples are then generated by reversing diffusion in probability space with the force data as a conditioner. The pLM is then used to decode the sequence for the final prediction. The generated sequences were tested by obtaining the three-dimensional structure via OmegaFold \cite{wu2022high} and then simulating the mechanical unfolding. It was shown that the generated sequences do produce responses close to the conditioned mechanical response. Another example is CPDiffusion, which is a DDPM treating sequences as categorical data embedded in Euclidean space \cite{zhou2023conditional}. In the study, the authors use CPDiffusion to generate novel programmable endonucleases. The designed proteins (after an initial screening) were validated experimentally, with all 27 being soluble. Furthermore, some of the proteins demonstrated desired properties, with 24 of them displaying DNA cleavage activity and most having enhanced enzymatic activity compared to the template protein.

Another novel diffusion method was used in the development on LatentDiff \cite{fu_latent_2023}. As the name suggests, diffusion is not carried out in coordinate space but rather in a latent space. Inspired by image generation methods like StableDiffusion \cite{rombach_high-resolution_2022}, Fu et al.\ create an $\se3$ autoencoder to downsample the geometry of the protein as well as upsample to recover the protein coordinates. Protein generation is carried out by reversing diffusion in the latent space and then upsampling to retrieve the predicted protein. In benchmarking, the method outperforms other small-protein generation methods, such as FoldingDiff, in designability with comparable (but slightly worse) results to FrameDiff. However, its sampling speed is around 64 times faster than FrameDiff, likely due to the reduced dimensionality of the latent space.

Protpardelle differs from the backbone generation methods in that it generates proteins at an all-atom resolution \cite{chu_all-atom_2023}. An obstacle to all-atom generation is that the number and types of atoms are determined exactly by the amino acid sequence, so making a choice on the coordinates to diffuse will also determine the amino acid sequence and thus the final structure of the protein. To address this, diffusion is carried out on a ``superposition'' of possible structures.  That is, the positions of atoms for any choice of amino acid are tracked during diffusion, allowing the model the flexibility to decide the sequence during the generation process. On the standard benchmarks, Protpardelle does quite well. In particular, its scRMSD score is about 90\% for small proteins and around 70\% for proteins around 300 amino acids in length. Also, because the model generates an all-atom representation, it is possible to condition generation on side-chain information. However, improvements in conditional generation are needed as the model is prone to ignoring the conditioning information.

While an overwhelming amount of work has been done on the generation and design of proteins, relatively less attention has been paid to other biomolecules. However, there have been some recent efforts to bring diffusion methods to DNA and RNA generation.

For instance, DiscDiff is a latent diffusion model for generating DNA sequences \cite{li_latent_2023}. A variational autoencoder is trained to first take DNA sequences into a latent space. Then a diffusion model is trained in latent space to produce samples. One notable change that the authors make is during the decoding step. Often with latent diffusion models, errors may be introduced when moving from latent space to sequence space. The authors hypothesize that the latent variable captures the global information of the sequence well but that there may be regions in the sequence with lower certainty. To address the authors introduce the Absorb-Escape method, where a pretrained autoregressive model for DNA sequences is used to correct nucleotides with lower certainty. When benchmarking against other sequence generation methods trained on the DNA task, it achieved state-of-the-art results without post-conditioning and did even better with Absorb-Escape.

Another attempt of bringing diffusion models to other biomolecules can be seen with MMDiff \cite{morehead_towards_2023}. This model jointly generates sequences and structures of nucleic acid and protein complexes. Many of the methods are directly inspired by the work of diffusion models applied to proteins. The diffusion takes place in \(\se3\) with coordinate frames for each of the residues of the protein and RNA/DNA structures, similar to that used by Yim et al.\ \cite{yim_se3_2023}. Sequences are simultaneously generated by embedding them into continuous space via a one-hot encoding, as done by Lisanza et al.\ for Protein Generator \cite{lisanza_joint_2023}. Benchmarking proceeds analogously to protein generation, where designability, diversity, and novelty are assessed. However, MMDiff fails to achieve high designability. For instance, the percentage of designable protein-nucleic acid complexes is 0.74\% (versus 0.37\% for random generation). The method does marginally better with the generation of nucleic acids only, with 8.67\% being designable (1.33\% for random generation). The authors suggest that the key to improvement is more diverse and high-quality training data. This highlights the current challenges in directly transferring structural generation models to the problem of RNA and DNA generation. 

\subsubsection{Molecular dynamics and ensemble generation}

Molecular dynamics simulations is an important tool for computational biology. It sees uses in the refinement of predictions and the study of dynamics of biomolecules \cite{feig2016protein,feig2017computational,heo2018prefmd,heo2018experimental,mirjalili2013protein,mirjalili2014physics,lindorff2011fast}. Typically, simulations are carried out using Langevin dynamics and a prescribed potential energy function. However, computation of a conventional energy function is expensive (usually quadratic in the number of particles), and so simulating the dynamics on a suitably long time-scale is infeasible. Generative models have the ability to speed up these computations by directly sampling the distribution of future states conditioned on the current state. This has been applied to the simulation of molecules with diffusion models and other generative models  \cite{wu_diffmd_2023,hsu_score_2023}. However, this direction has not been thoroughly explored for the simulation of protein dynamics.

Lu et al.\ made progress in this direction with the introduction of Str2Str, a model for sampling protein conformations \cite{lu_str2str_2023}. The method shares similarities to simulated annealing, where the temperature for MD simulations is raised to overcome energetic barriers. Here a structure is perturbed by diffusion, and then the process is reversed to sample a new conformation of the protein. The noising process is carried out in \(\se3\) as described by Yim et al.\ \cite{yim_se3_2023} and the score network is given by a variant of IPA \cite{jumper_highly_2021}. Str2Str is able to recapture many of the statistics of the equilibrium distributions of fast-folding proteins in negligible amount of time compared to direct MD simulation (510 GPU seconds versus over 160 GPU days, respectively.)

A similar approach was carried out by Zheng et al.\ with the development of the Distributional Graphormer (DiG) \cite{zheng2024predicting}. This model samples the equilibrium distribution of molecular systems using a diffusion model, where the score architecture is given by a Graphormer. In this work, the equilibrium distribution of proteins is used during training and validation. Again, diffusion is carried out over frame data in \(\mathrm{SE}(3)\) using the VE formulation of the SDE. One notable addition to the typical diffusion model is the use of physics-informed diffusion pre-training (PIDP). Since MD simulation are expensive to compute, the authors sought a method to utilize prior energy functions to train the score model. The main idea is to enforce that the score network satisfies the corresponding Fokker-Plank equation -- a partial differential equation which governs the value of the score through the diffusion process -- and the initial condition given by the gradient of the prior energy. The PIDP does not require samples from the equilibrium distribution since information from the energy function will enforce the correct distributions. A tractable number of samples can be selected during this process since the equilibrium distribution lies on a much lower dimensional manifold than the latent space. DiG is shown to recapture protein equilibirum distributions, as well as generate plausible transition paths between conformations.

A related direction is the training of coarse-grained models for proteins. The free-energy landscape for an all-atom representation of a protein is rough and difficult to simulate, so coarse-grained models look to simplify the simulations by combining particles. The training of coarse-grained models is typically done through variational force matching, where the forces on the virtual particles are chosen to match the expected force before the coarse-graining process. However, this process is data-inefficient since the coarse-graining maps many configurations to the same state and so forces must be averaged out over those configurations. An alternative method is using relative entropy, but this requires resampling the coarse-grained distribution during training. Flow-matching is a more data-efficient method of training a coarse-grained model \cite{kohler_flow-matching_2023}, which first uses a generative model to learn the equilibrium distribution of a protein that then is used to determine the forces of the model. This method is built on by Arts et al.\, who train a generative and coarse-grained model simultaneously using a denoising diffusion generative model \cite{arts2023two}. Similarly to flow matching, training does not require the recording of forces. The coarse-grained model and sampler are able to recapture the equilibrium distribution of the proteins on which they are trained. The training scheme shows promise in terms of scaling to longer proteins.

\subsection{Predictive modeling}

Diffusion models being generative models are traditionally designed to sample the whole probability distribution they are trained on. However, there has been success with using diffusion models to predict the solution to regression and/or optimization problems with a single solution. Non-generative machine learning methods sometimes struggle to account for uncertainty and thus try to minimize the expected error at the cost of realistic predictions (see \cite[\S 3]{corso_diffdock_2023} for an example). A generative model instead handles the uncertainty by sampling over all possibilities, giving a more accurate representation of the likely solution. 

For generating a single prediction, one can repeatedly sample from the diffusion model and then choose the ``best'' sample. A typical strategy for making the final choice is training a separate confidence model which would rank each of the samples based on the confidence of being the optimal solution. A training-free alternative would be to use a likelihood estimate or evidence-based lower bound (ELBO) to determine the most likely sample from the distribution.

Following this strategy, several diffusion models have been applied to prediction problems and have achieved state-of-the-art results.

\subsubsection{Protein side-chains}

The protein side-chain packing problem focuses on the case where the backbone of a protein is fixed and asks for the positions of its side-chain atoms. Traditional methods for solving the problem involve searching the space of positions for the side-chains using rotamer libraries and comparing energies \cite{huang2020faspr,yanover2008minimizing,liang2011fast,badaczewska2020computational,alford2017rosetta,krivov2009improved,xu2006fast,cao2011improved}. Due to the complexity of the energy landscape and the number of possible conformations of the side-chains, this process can be time-intensive. Machine-learning methods aim to speed up this process while maintaining precision. Methods such as AttnPacker \cite{mcpartlon_end--end_2022} predict the positions of the side-chain atoms but do not account for bond lengths and angles. Other machine learning methods treat side-chain packing as a regression problem and do not model the diversity of possible conformations in the energy landscape \cite{misiura2022dlpacker,chaudhury2010pyrosetta}.

DiffPack is a diffusion-based model for side-chain packing that predicts the torsional angles, \(\chi_1, \chi_2, \ldots\), of each side-chain \cite{zhang_diffpack_2023}. Ideal bond lengths and angles are assumed for the side-chain atoms, so that the degrees of freedom are greatly reduced and the samples produced have natural bond lengths and angles. The diffusion process is carried out on \(\mathbb T^n\), where a sample corresponds to a choice of torsional angles for the side-chains. One notable change in the diffusion process is that the angles are sampled autoregressively: \(\chi_1\) is first sampled, followed by \(\chi_2\) and so on. Since the value of \(\chi_1\) affects the coordinates of atoms further down the chain, sampling all angles simultaneously can lead to excess steric clashes due to the sensitivity of earlier angles on coordinates. Using an autoregressive generation process mitigates this effect because the model can self-correct for earlier choices of angles. A confidence model is trained along side the diffusion model in order to score the sampled conformations. During inference, multi-round sampling and annealed temperature sampling are used to choose conformations with lower energies. DiffPack is able to achieve state-of-the-art performance in the mean absolute error of the predicted torsional angles when compared to other machine learning approaches to the side-chain packing problem. Diffpack's reported results are given in \cref{tab:casp13,tab:casp14}. For details on the benchmarking, see \cref{app:diffpack}.

\begin{table}[ht]
	\centering
	\begin{adjustbox}{max width=\linewidth}
		\begin{tabular}{lcccc|ccc|ccc}
			\toprule
			& \multicolumn{4}{c}{\textsc{Angle MAE $^{\circ}$} $\downarrow$} & \multicolumn{3}{c}{\textsc{Angle Accuracy \%} $\uparrow$} & \multicolumn{3}{c}{\textsc{Atom RMSD Å} $\downarrow$} \\
			\cmidrule(lr){ 2 - 5 } \cmidrule(lr){ 6 - 8 } \cmidrule(lr){ 9 - 11 } \textbf{Method} & $\chi_1$ & $\chi_2$ & $\chi_3$ & $\chi_4$ & All & Core & Surface & All & Core & Surface \\
			\midrule
			SCWRL           & 27.64 & 28.97 & 49.75 & 61.54 &  56.2\% &  71.3\% & 43.4\% & 0.934 & 0.495 & 1.027\\
			FASPR           & 27.04 & 28.41 & 50.30 & 60.89 &  56.4\% &  70.3\% & 43.6\% & 0.910 & 0.502 & 1.002 \\
			RosettaPacker   & 25.88 & 28.25 & 48.13 & 59.82 &  58.6\% &  75.3\% & 35.7\% & 0.872 & 0.422 & 1.001 \\
			DLPacker        & 22.18 & 27.00 & 51.22 & 70.04 &  58.8\% &  73.9\% & 45.4\% & 0.772 & 0.402 & 0.876\\
			AttnPacker      & 18.92 & 23.17 & 44.89 & 58.98 &  62.1\% &  73.7\% & 47.6\%  & 0.669 & 0.366 & 0.775 \\
			\midrule
			% DiffPack        & \textbf{15.99} & \textbf{19.64} & \textbf{38.85} & \textbf{50.82}  &  \textbf{69.5\%} &  \textbf{82.7\%} & \textbf{57.3\%}  & \textbf{0.597} & \textbf{0.298} & \textbf{0.696}\\
			DiffPack        & \textbf{15.35} & \textbf{19.19} & \textbf{37.30} & \textbf{50.19}  &  \textbf{69.5\%} &  \textbf{82.7\%} & \textbf{57.3\%}  & \textbf{0.579} & \textbf{0.298} & \textbf{0.696}\\
			\bottomrule
		\end{tabular}
	\end{adjustbox}
	\vspace{3pt}
	\caption{Comparative evaluation of DiffPack and prior methods on CASP13 as reported in \cite{zhang_diffpack_2023}.}
	\label{tab:casp13}
\end{table}
\begin{table}[ht]
	\centering
	\begin{adjustbox}{max width=\linewidth}
		\begin{tabular}{lcccc|ccc|ccc}
			\toprule
			& \multicolumn{4}{c}{\textsc{Angle MAE $^{\circ}$} $\downarrow$} & \multicolumn{3}{c}{\textsc{Angle Accuracy \%} $\uparrow$} & \multicolumn{3}{c}{\textsc{Atom RMSD Å} $\downarrow$} \\
			% \cline { 2 - 5 } \cline { 9 - 11 } & $\chi_1$ & $\chi_2$ & $\chi_3$ & $\chi_4$ & & All & Core & Surface \\
			\cmidrule(lr){ 2 - 5 } \cmidrule(lr){ 6 - 8 } \cmidrule(lr){ 9 - 11 } \textbf{Method} & $\chi_1$ & $\chi_2$ & $\chi_3$ & $\chi_4$ & All & Core & Surface & All & Core & Surface \\
			\midrule
			SCWRL           & 33.50 & 33.05 & 51.61 & 55.28             &  45.4\%   & 62.5\%  & 33.2\% & 1.062 & 0.567 & 1.216\\
			FASPR           & 33.04 & 32.49 & 50.15 & 54.82             &  46.3\%   & 62.4\%  & 34.0\% & 1.048 & 0.594 & 1.205 \\
			RosettaPacker   & 31.79 & 28.25 & 50.54 & 56.16             &  47.5\%   & 67.2\%  & 33.5\% & 1.006 & 0.501 & 1.183\\
			DLPacker        & 29.01 & 33.00 & 53.98 & 72.88             &  48.0\%   & 66.9\%  & 33.9\% & 0.929 & 0.476 & 1.107 \\
			AttnPacker      & 25.34 & 28.19 & 48.77 & \textbf{51.92}    &  50.9\%   & 66.2\%    & 36.3\%    & 0.823 & 0.438 & 1.001  \\
			\midrule
			% DiffPack        & \textbf{23.43} & \textbf{26.04} & \textbf{45.41} & 56.64  &  \textbf{57.5\%} & \textbf{77.8\%} & \textbf{43.5\%} & \textbf{0.793} &\textbf{0.356} & \textbf{0.956} \\
			DiffPack        & \textbf{21.91} & \textbf{25.54} & \textbf{44.27} & 55.03  &  \textbf{57.5\%} & \textbf{77.8\%} & \textbf{43.5\%} & \textbf{0.770} &\textbf{0.356} & \textbf{0.956} \\
			\bottomrule
		\end{tabular}
	\end{adjustbox}
	\vspace{3pt}
	\caption{Comparative evaluation of DiffPack and prior methods on CASP14 as reported in \cite{zhang_diffpack_2023}.}
	\label{tab:casp14}
\end{table}

A related problem to side-chain packing for \textit{apo} protein structures is determining the conformations of side-chains for ligand-docking problems. The native structure of proteins do not account for the the complex flexibility of proteins, which adapt their structure to their respective molecular binders. Thus there is demand for methods which take this flexibility into account. PackDock seeks to incorporate this flexibility to improve on ligand-protein docking predictions \cite{zhang2024packdock}. The authors note that the dynamics of the protein-ligand process are commonly described by two mechanisms: conformational selection and induced fit. A torsional diffusion method -- called PackPocket --  is introduced to sample multiple possible conformations for the side-chains, either with or without the ligand. This is carried out in conjunction with a docking algorithm to sample ligand conformations in the binding pocket (Vina \cite{trott2010autodock} is used by the authors, but any docking method can be used). PackDock was shown to improve docking predictions when considering \textit{apo} structures or \textit{holo} structures form non-homologous ligands.

\subsubsection{Protein structure prediction}

Diffusion models have also been applied to the problem of predicting the native structure of proteins. In particular, EigenFold predicts the coordinates of the C\(\alpha\) atoms of a protein \cite{jing_eigenfold_2023}. Diffusion is carried out on the Cartesian space \(\mathbb R^{3n}\), but the diffusion process is defined so that the prior distribution uses a Harmonic potential for the relative distances of points. This guarantees that diffused points do not separate too far and that relative distances more closely reflect observed bond lengths. Thus the prior distribution models more realistic molecular conformations, but the stiffness of the diffusion process requires the SDE be carried out with a basis of eigenvectors. An estimated ELBO is used \emph{in lieu} of training a score model. The results reported in \cref{tab:cameo} are modest compared to established machine learning approaches such as AlphaFold2 and ESMFold \cite{jumper_highly_2021,lin2023evolutionary} (see \cref{app:eigenfold} for details). One benefit to using a generative model over a single-structure prediction methods is that generative models can naturally capture conformational diversity of a protein. If a protein is flexible, then sampling the distribution of conformations should result in different possible poses. Eigenfold shows some moderate results in this direction, with a slight correlation between the diversity of samples and the flexibility of the protein. Overall, the model demonstrates the possibilities of diffusion models to the problem of predicting native structures.

\begin{table}[hb]
	\centering
	\begin{tabular}{lcccc}
		\toprule 
		& RMSD$_{\text{C}\alpha}$ $\downarrow$ & TMScore $\uparrow$& GDT-TS $\uparrow$& lDDT$_{\text{C}\alpha}$ $\uparrow$\\ \midrule
		\textsc{AlphaFold2} & 3.30 / 1.64 & 0.87 / 0.95  & 0.86 / 0.91 & 0.90 / 0.93 \\
		\textsc{ESMFold} & 3.99 / 2.03 & 0.85 / 0.93 & 0.83 / 0.88 & 0.87 / 0.90\\
		\textsc{OmegaFold} & 5.26 / 2.62 & 0.80 / 0.89 & 0.77 / 0.84 & 0.83 / 0.89 \\
		\textsc{RoseTTAFold} & 5.72 / 3.17 & 0.77 / 0.84 & 0.71 / 0.75 & 0.79 / 0.82 \\
		\midrule
		\textsc{EigenFold} & 7.37 / 3.50 & 0.75 / 0.84 & 0.71 / 0.79 & 0.78 / 0.85\\
		\bottomrule
	\end{tabular}
	\caption{Single-structure prediction accuracy of \textsc{EigenFold} and baseline methods on CAMEO targets under 750 residues from Aug 1--Oct 31, 2022. All metrics are reported as mean / median. Table from \cite{jing_eigenfold_2023}}
	\label{tab:cameo}
\end{table}

A recent development in protein-structure prediction was the announcement of AlphaFold 3 (AF3) \cite{abramson2024accurate}. The main improvement of AF3 over AF2 \cite{jumper_highly_2021} and AlphaFold-Multimer \cite{evans2021protein} is its capability of modeling general molecules. In particular, the accelerated review paper cites state-of-the-art results in ligand-protein complex predictions and advancements in protein-RNA modeling compared to other automated methods. To accommodate the more general modeling of biomolecular interactions, the generation process was switched to an all-atom framework that can generalize the construction process of proteins and other molecules that do not have amino acid nor nucleic acid residues. Since frames and invariant point attention cannot be used in this case, the structure module of AF2 was changed to a diffusion module in AF3. The diffusion module takes the embedding information created by AF3 in order to determine the atom positions of the proteins and other molecules, where the generation process follows a standard iterative process of removing Gaussian noise from a sample. One notable difference between AF3 and other applications of diffusion to biomolecules is the lack of invariance/equivariance to rotations and translations in the architecture. Instead, data augmentation is used to achieve equivariance.

A related but narrower problem is that of backmapping: the task of determining an all-atom configuration from a coarse-grained representation. Data-driven approaches to the problem seek to accurately reconstruct the protein while avoiding un-physical results such as clashes. While some methods are deterministic, an additional goal would be to capture the diversity of different possible all-atom configurations in a thermodynamically consistent way. Previous state-of-the-art methods, used generative models such as GANs \cite{goodfellow2020generative} and variational auto-encoders (VAEs) \cite{kingma2013auto}. In particular, GenZProt uses a VAE to backmap a C$\alpha$ trace to a full-atom resolution while still generalizing to molecules outside the training set. Researchers have begun to use diffusion models to generate backmapping samples. DiAMoNDBack, for instance, uses diffusion to backmap C$\alpha$ traces \cite{jones2023diamondback}. The backmapping is done residue-by-residue in an autoregressive fashion, which lends to a more transferable model. Another approach to this problem is BackDiff, which more generally backmaps any coarse-grained representation \cite{liu2023backdiff}. Determining the position of heavy-atoms given the coarse-grained atoms is treated as an imputation problem, and training is done by randomly choosing heavy atoms (apart from the C$\alpha$ atoms) to mask. Conditioning on auxiliary variables and enforcing bond lengths and angles are handled during generation using a manifold constraint. Both DiAMoNDBack and BackDiff are shown to outperform GenZProt in terms of accuracy of full-atom reconstructions, diversity of samples, and the avoidance of clashes and un-physical bond lengths and angles.

\subsubsection{Inverse protein folding}

As opposed to structural prediction methods, the inverse folding problem looks at determining an amino acid sequence given a protein's C\(\alpha\) backbone. GraDe-IF applies a discrete diffusion model to the inverse folding problem \cite{yi_graph_2023-1}. The diffusion is carried out in the categorical space of amino acid types using the D3PM framework \cite{austin_structured_2023}. Notably, the authors use the Blocks Substitution Matrix (BLOSUM) \cite{henikoff1992amino} to create the transition matrix for the noising step. This allows the model to explore closely related amino acids during the latter stage of the denoising process instead of random, unrelated residue types. The score network is given by an adaptation of an equivariant graph neural network (EGNN) \cite{satorras_en_2021} to account for the roto-translational invariance of the problem. The model outperforms other inverse folding methods in the authors' benchmarking as reported in \cref{tab:rr}. Details of the benchmarking can be found in \cref{app:gradeif}.

\begin{table}[!t]
	\begin{center}
		\resizebox{\textwidth}{!}{
			\begin{tabular}{ccccccccc}
				\toprule
				\multirow{2}{*}{\textbf{Model}} & \multicolumn{3}{c}{ \textbf{Perplexity} $\downarrow$} & \multicolumn{3}{c}{ \textbf{Recovery Rate} \% $\uparrow$} & \multicolumn{2}{c}{\textbf{CATH version}} \\\cmidrule(lr){2-4}\cmidrule(lr){5-7}\cmidrule(lr){8-9}
				& Short & Single-chain & All & Short & Single-chain & All & 4.2 & 4.3 \\
				\midrule 
				\textsc{StructGNN} & 8.29 & 8.74 & 6.40 & 29.44 & 28.26 & 35.91 & $\checkmark$ & \\
				\textsc{GraphTrans} & 8.39 & 8.83 & 6.63 & 28.14 & 28.46 & 35.82 & $\checkmark$ & \\
				GCA  & 7.09 & 7.49 & 6.05 & 32.62 & 31.10 & 37.64 & $\checkmark$ & \\
				GVP  & 7.23 & 7.84 & 5.36 & 30.60 & 28.95 & 39.47 & $\checkmark$ & \\
				GVP-large & 7.68 & 6.12 & 6.17 & 32.6 & 39.4 & 39.2 & & $\checkmark$ \\
				\textsc{AlphaDesign} & 7.32 & 7.63 & 6.30 & 34.16 & 32.66 & 41.31 & $\checkmark$ & \\
				\textsc{ESM-if1}  & 8.18 & 6.33 & 6.44 & 31.3 & 38.5 & 38.3 & & $\checkmark$ \\
				\textsc{ProteinMPNN}  & 6.21 & 6.68 & 4.57 & 36.35 & 34.43 & 49.87 & $\checkmark$ & \\
				\textsc{PiFold} & $6.04$ & 6.31 & $4 . 5 5$ & $39.84$ & 38.53 & 51.66 & $\checkmark$ & \\
				\midrule
				\textsc{GraDe-IF} &$\mathbf{5.49}$&$\mathbf{6.21}$&$\mathbf{4.35}$&$\mathbf{45.27}$&$\mathbf{42.77}$&$\mathbf{52.21}$&$\checkmark$\\
				\bottomrule
			\end{tabular} 
		}
	\end{center}
	\caption{Recovery rate performance of \textbf{CATH} on zero-shot models. Table taken from \cite{yi_graph_2023-1}}
	\label{tab:rr}
\end{table}

\subsubsection{Protein complexes}

Ligands bind to proteins and modulate their biological function, so an important step in drug design is determining where a ligand will bind to a protein. Traditionally, computational methods for this problem search for the ligand position that directly optimizes a score function \cite{trott2010autodock,friesner2004glide,thomsen2006moldock}. However, due to the high-dimensionality of the space and the roughness of the energy landscape, this process can be slow and give inaccurate results. Machine learning approaches that treat ligand-binding as a regression problem significantly speed up the process but still suffer from inaccuracy due to the uncertainty in the energy landscape \cite{stark2022equibind,lu2022tankbind}. DiffDock \cite{corso_diffdock_2023} instead treats the finding the ligand pose as a generative problem. Put simply, DiffDock samples possible ligand poses from an underlying distribution and then makes an optimal choice based on a learned confidence score. This circumvents the ruggedness of the energy landscape by enumerating multiple options instead of averaging possibilities together as done in regression-type approaches. DiffDock uses a diffusion model to sample different poses and a separate score model to determine the confidence that the RMSD is below 2 \r{A}. The dimensionality of the search space is reduced by fixing the conformation of the protein and assuming ideal bond lengths and bond angles in the ligand. Thus the degrees of freedom are the position, orientation, and dihedral angles of the ligand. Then sampling a ligand pose is the same as sampling from the manifold \(\mathbb R^3 \times \so3 \times \mathbb T^m\), where \(m\) is the number of torsional angles. DiffDock was benchmarked on the PDBBind dataset \cite{liu2017forging} with a time split at 2019 for training and validation. On this data set, DiffDock reported state-of-the-art performance, and it retained high accuracy on computationally obtained structures. FusionDock augments DiffDock's approach by using physics-informed priors \cite{masters_fusiondock_nodate}.

However, the reported results tell an incomplete story of DiffDock's strength. Some noted that the comparison to classical docking methods is unfair since traditional methods were not designed for blind docking and typically require a designated pocket to perform well \cite{yu2023deep}. When given a binding pocket, traditional methods outperform deep learning methods. But experiments demonstrated that DiffDock still excels at pocket-finding. Others have pointed out that the typical time split in PDBBind can lead to data leakage in the validation; some of the structures in the training set are similar to those in the validation set, leading to an overestimation of a model's performance \cite{kanakala2023latent,li2023leak}. Finally, many of the predictions of deep learning docking methods are physically implausible, and when accounting for docking predictions that are within 2 Å of the true structure \emph{and} realistic, deep learning methods perform substantially worse than classical methods \cite{buttenschoen2024posebusters}. Recently, Corso et al.\ created a new benchmark set, DockGen, that removes the data leakage in terms of structure and pocket similarlity \cite{corso2024deep}. In addition, they introduce DiffDock-L, a larger model trained on a bigger, augmented data set. On both the PDBBind and DockGen benchmarks, DiffDock-L outperforms all other blind docking methods. \Cref{tab:results_main} details the results of DiffDock-L. More information on benchmarking is given in \cref{app:diffdock}.

\begin{table}[ht]
	\begin{small}
		\begin{center}
			
			\begin{tabular}{lcc|cc|cc|c}
				\toprule
				& \multicolumn{2}{c|}{PDBBind}                            & \multicolumn{2}{c|}{\textsc{DockGen}-full}               & \multicolumn{2}{c|}{\textsc{DockGen}-clusters}           & Average               \\
				Method                                                            & \%$<$2\AA{}         & Med. & \%$<$2\AA{} & Med. & \%$<$2\AA{} & Med. & Runtime (s)           \\ \midrule
				\textsc{SMINA}                   & 18.7           & 7.1   &  7.9      & 13.8       &  2.4      &  16.4      & 126*  \\ 
				\textsc{SMINA (ex. 64)}          & 25.4           & 5.5   &  10.6      &  13.5      &  4.7      & 14.7       & 347*  \\
				\textsc{P2Rank+SMINA}            & 20.4           & 4.3   &  7.9      & 14.1       &  1.2      &  16.4      & 126*  \\
				\textsc{GNINA}                   & 22.9           & 7.7   &  14.3     & 15.2       &  9.4      &  14.5      & 127   \\
				\textsc{GNINA (ex. 64)}          & 32.1           &  4.2  &  17.5      & 8.1       &   11.8     & 6.2       & 348   \\
				\textsc{P2Rank+GNINA}            & 28.8           & 4.9   &  13.8     & 16.2       &   4.7     &  15.3      & 127   \\ \midrule
				\textsc{EquiBind}                & 5.5            & 6.2   &  0.0      & 13.3       &  0.0      & 13.3       & \textbf{0.04}  \\
				\textsc{TANKBind}                & 20.4           & 4.0   & 0.5   & 11.6       & 0.0       & 11.1       & 0.7               \\
				\textsc{DiffDock} (10)           & 35.0           & 3.6   & 7.1       & 6.8       &  6.1      & 6.0       & 10    \\
				\textsc{DiffDock} (40)           & 38.2  &  3.3  & 6.0       & 7.3       &  3.7      & 6.7       & 40    \\ \midrule \midrule
				\textsc{DiffDock-L} (10)           & \textbf{43.0}  & \textbf{2.8}   & \textbf{22.6}       & \textbf{4.3}       &  \textbf{27.6}      &  \textbf{3.7}      &  25  
			\end{tabular}
		\end{center}
	\end{small}
	\caption{ Top-1 RMSD performance of different methods on the PDBBind and \textsc{DockGen} benchmarks. Runtimes were computed as averages over the PDBBind test set. * run on CPU. Med. indicates the median RMSD. Table from \cite{corso2024deep}.}
	\label{tab:results_main}
\end{table}

A similar framework was applied to the protein-protein docking problem with DiffDock-PP \cite{ketata_diffdock-pp_2023}. The diffusion process from DiffDock is adapted by assuming that the proteins are rigid bodies so that diffusion takes place on \(\mathbb R^3 \times \so3\) with torsional fixed. Benchmarking showed improvements in accuracy compared to other machine learning methods while being much faster than traditional search methods when computed on GPU.

NeuralPLexer is another diffusion-based method that predicts ligand-protein docking complexes \cite{qiao_state-specific_2023}. Unlike previous docking prediction models, this method does not rely on the protein structure as an input but only takes the protein sequence and ligand molecular graph. NeuralPLexer is comprised of a graph-based architecture to encode molecular properties, a contact prediction module that determines the inter-molecular distances, and an equivariant diffusion module that determines the structure of the complex at an atomic resolution. When applied to the blind ligand-protein docking problem (a simplification of the full complex prediction), NeuralPLexer outperformed other methods like DiffDock.The model also outperforms AlphaFold2 on protein structure prediction for structures with high flexibility and newly added protein-ligand complexes.\footnote{A recent press release claims improved performance for the newly trained NeuralPLexer2, although details have yet to be released.} 

A similar model to NeuralPLexer is the Diffusion model for Protein-Ligand complexes (DPL) \cite{nakata_end--end_2023}. Like NeuralPLexer, it does not take protein structure as an input, but only relies on the protein sequence and ligand molecular graph. Furthermore, DPL does not rely on protein backbone templates during prediction. The inputs are embedded into a single and pair representation with information from a protein language model (ESM-2) \cite{verkuil_language_2022} and these representations are updated with Folding blocks from ESMFold \cite{lin2023evolutionary}. Finally, and equivariant denoising network is used to generate the complex. DPL is able to generate diverse sets of complexes, but performance is limited for complexes with less data.

%% file: sections/conclusion.tex
\section{Conclusion}

Diffusion models have made remarkable progress in the design and prediction of proteins, with many recent models quickly achieving state-of-the-art performance on tasks. Protein backbone generation methods, such as RFdiffusion, have demonstrated incredible success at sampling realistic and novel structures that have been validated experimentally \cite{watson_novo_2023}. In addition, the ability to conditionally generate structures opens up new avenues for designing proteins with desired properties. Diffusion models have also seen great success with other generation tasks, such as sequence generation/design and ensemble sampling.

They have also seen success in predictive tasks. Generative models are better at sampling many possible modes in a distribution (as opposed to regression models which average over possibilities leading to unrealistic results), which gives them an edge in some predictive tasks. Diffusion models in particular seem well suited to searching a large-dimensional, continuous space, as partially evidenced by DiffDock's ability to find pockets for protein-ligand docking \cite{yu2023deep}.

One direction that we believe remains underexplored is the use of diffusion models for generating and predicting DNA/RNA structures. As seen with the results with proteins, diffusion models provide a potentially potent method for solving problems in this domain. Attempts at structure prediction has led to mild results \cite{morehead_towards_2023}. Some success has been achieved with jointly modeling them with proteins (\cite{abramson2024accurate}, for instance) where deep knowledge of protein structure can be leveraged, but there has been little success in monomeric prediction of these biomolecules alone. To this end, there are some challenges to directly applying diffusion methods: (1) there is relatively less data for nucleic acids as compared to proteins, which makes training more difficult; and (2) RNA structures are more flexible, which mirrors the difficulty with certain protein generation methods from creating intrinsically disordered regions.

It seems certain that in the coming years there will be more attention paid to diffusion models for applications involving biomolecules, but they are not without competitors in the space. Inspired by the success of diffusion, researchers have proposed other generative deep learning methods that take an iterative approach to generation. For instance, Bayesian flow networks (BFNs)  are generative models which iteratively samples noised distributions in a similar way to diffusion models \cite{graves2023bayesian}. The sampling process consists of continually applying Bayes' rule to flow toward a desired distribution to sample. BFNs naturally lend themselves to multimodal problems and have been used in molecular generation \cite{song2023unified}. A more popular approach to generative modeling has been flow matching \cite{lipman2022flow}. Flow matching is a subset of continuous normalizing flows where the diffeomorphism between probability distributions is given by the pushforward of a learned vector field that is trained with conditional flow matching. Sampling then proceeds by solving the differential equation given by the vector field. An advantage of this method over diffusion models is reduced training and sampling time. Flow matching has seen success in backbone generation \cite{bose2023se,yim2023fast} and sampling protein ensembles \cite{jing2024alphafold}. 

%^maybe include paper on motif scaffolding to highlight conditional generation
Regardless of the exact shape of the field in the coming years, it is undeniable that diffusion models and similar frameworks will continue to be used in the design and prediction of biomolecules. They have introduced scalable, generative modeling to the field and have already demonstrated their advantages over previous traditional and machine-learning approaches by setting new state-of-the-art results in numerous problems.

%% file: sections/acknowledgments.tex
\section{Acknowledgments}
This work was partially supported by the National Institute of General Medical Sciences [R35GM138146 to D.B.] and the National Science Foundation [DBI2208679 to D.B.].

%% file: sections/benchmarking.tex
\section{Benchmarking details}

\subsection{Backbone generation methods}
\label{app:backbone}

We focus on methods that were able to generate proteins up to 250 amino acids long. We generated 50 proteins each for lengths 50, 100, 150, 200, and 250 (giving a total of 250 generated samples). To test designability, we used ProteinMPNN \cite{dauparas2022robust} at a temperature of 0.1 to create 8 possible sequences, which were then folded using ESMFold \cite{lin2023evolutionary}. The minimum difference between the RMSD of the folded C\(\alpha\) backbones and the corresponding sample was recorded as the scRMSD for the structure. A protein is considered designable if scRMSD is less than 2.0 Å, and the percentage of designable proteins is reported for each method. To measure diversity, TM-align \cite{zhang2005tm} is used to compare each pair of designable proteins from a method (normalizing with the length of the longer protein). The average TM-Score is recorded for each method. To measure novelty, we used FoldSeek \cite{van2022foldseek} to compare the designable proteins against the PDB \cite{berman2003announcing}. The TM-Score is computed against proteins in the PDB and the maximum is recorded as PDBTM. If PDBTM is less than 0.5, then the generated structure is considered novel. The percentage of novel proteins is recorded.

\subsection{DiffPack}
\label{app:diffpack}

Training and validation used BC40 \cite{wang2016protein} and followed the dataset split given in AttnPacker \cite{mcpartlon_end--end_2022}. The models were then evaluated on CASP13 and CASP14 datasets. The training set was curated to remove sequences with 40\% or more similarity to those in the test set. Both deep learning methods, like AttnPacker \cite{mcpartlon_end--end_2022} and DLPacker \cite{misiura2022dlpacker}, and traditional methods, like SCWRL4 \cite{krivov2009improved}, FASPR \cite{huang2020faspr}, and RosettaPacker \cite{chaudhury2010pyrosetta}, are used as baselines for DiffPack. 

For each model the angle MAE, angle accuracy, and atom RMSD are computed. Angle MAE is the mean absolute error of the predicted side-chain torsional angles versus the actual angles. Angle accuracy is the percentage of angles that are within 20$\deg$ of the actual angle. Atom RMSD is the average RMSD of the side-chain atoms for each residue. Results are further categorized by ``Core" and ``Surface" residues. Residues are labeled as core residues if at least 20 $C_\beta$ atoms are within a 10 {\AA} radius. Surface residues have at most 15 $C_\beta$ atoms within the same radius.

\subsection{Eigenfold}
\label{app:eigenfold}

The test set in the Eigenfold paper \cite{jing_eigenfold_2023} is CAMEO targets released between August 1st, 2022 and October 31st, 2022 with lengths of less than 750 residues. The Eigenfold method was benchmarked against other structure prediction methods: namely, AlphaFold2 \cite{jumper_highly_2021}, ESMFold \cite{lin2023evolutionary}, OmegaFold \cite{wu2022high}, and RoseTTAFold \cite{baek2021accurate}. For each method, five protein structures were generated and ranked based on an approximate ELBO. The top-ranked structure was taken as the final prediction and compared against the native structure. The metrics in \cref{tab:cameo} are the results of these comparisons.

\subsection{GraDe-IF}
\label{app:gradeif}

The inverse folding methods were tested on recovering protein sequences in CATH \cite{orengo1997cath}. The training-validation-training split follows the CATH v4.2.0-based partitioning as done for GraphTrans \cite{ingraham2019generative} and GVP \cite{jing2020learning}. Testing was further split into three categories: short, single-chain, and all. \textit{Short} refers to proteins of length less than 100 residues. \textit{Single-chain} refers to proteins with only one chain. The methods were evaluated on \textit{perplexity} and \textit{recovery rate}. Perplexity  measures how well the predicted amino acid probabilities match the actual amino acids at each position. A lower perplexity means that the model better fits the distribution of the data. The recovery rate measures the percentage of amino acids that are correctly predicted from the original sequences. A higher recovery rate indicates the model is better at predicting the original sequence. In addition, the training data set for each model is given in \cref{tab:rr}: CATH version 4.2 or 4.3.

GraDe-IF was benchmarked against other inverse folding methods, including StructGNN \cite{ingraham2019generative}, GraphTrans \cite{ingraham2019generative}, GCA \cite{tan2022generative}, GVP \cite{jing2020learning}, GVP-large \cite{hsu2022learning}, AlphaDesign \cite{gao2022alphadesign}, ESM-if1 \cite{hsu2022learning}, ProteinMPNN \cite{dauparas2022robust}, and PiFold \cite{gao2022pifold}.

\subsection{DiffDock-L}
\label{app:diffdock}

The benchmarking in \cite{corso2024deep} was performed on the PDBBind dataset \cite{liu2017forging} and the authors' new dataset, DockGen. The main aim of the dataset was to provide better benchmarks for the protein-ligand docking problem by preventing data leakage between training and testing data. For instance, proteins with less than 30\% sequence similarity may nonetheless share very similar binding pockets, and so benchmarking must account for these structural similarities when doing a split of training and testing. The authors use the ECOD \cite{cheng2014ecod} classification to determine the protein domain of the chain making the most contacts with the ligand. This is used to cluster the complexes with an equal number of clusters being assigned to the training and testing datasets. Further filtering steps are taken (such as removing protein-ligand complexes with multiple ligands binding to the same pocket) to reach the final benchmark dataset. 

The benchmarking was carried out on both search-based and ML methods. The search-based methods included SMINA \cite{kohler_flow-matching_2023} and GNINA \cite{mcnutt2021gnina}. Since search-based methods have improved blind docking results by first selecting a pocket with a pocket finder method, results when using P2Rank \cite{krivak2018p2rank} prior to docking are also recorded. The ML methods include Equibind \cite{stark2022equibind}, TankBind \cite{lu2022tankbind}, and DiffDock \cite{corso_diffdock_2023}. For the DiffDock models, the number of poses sampled is also provided by the number in parentheses.